\documentclass{aa}

\usepackage{hyperref}
\usepackage{mathrsfs}
\usepackage{threeparttable}

\begin{document}

\title{Numerical simulations of the line-force-driven winds from active galactic nuclei: The special relativistic effects}
\titlerunning{Simulations of line-force-driven AGN winds}

\author{Luo-bin Tang\inst{1}
        \and Xiao-hong Yang \inst{1}\thanks{Corresponding author: Xiao-hong Yang}
        \and De-Fu Bu\inst{2}\thanks{Corresponding author: De-Fu Bu}
        }
\authorrunning{Tang, Yang, Bu}

\institute{Department of Physics and Chongqing Key Laboratory for Strongly Coupled Physics, Chongqing University, Chongqing 400044, People's Republic of China\\
            \email{yangxh@cqu.edu.cn}
            \and  Shanghai Key Lab for Astrophysics, Shanghai Normal University, 100 Guilin Road, Shanghai 200234, PR China\\
            \email{dfbu@shnu.edu.cn}}

\abstract
{}
   {Ultra-fast outflows (UFOs) with mildly relativistic velocities are frequently observed in active galactic nuclei (AGNs). The line-force-driving mechanism is often taken as a potential mechanism for driving UFOs. Due to the line-force-driven winds moving at mildly relativistic velocities, the special relativistic effects become important.}
   {There are two special relativistic effects: one is the influence of the disc rotation on the radiation field; the other is the radiation-drag effect. We wish to study the influence of the special relativistic effects on the line-force-driven winds, and we performed numerical simulations to investigate this.}
   {We find that the line-force-driven winds are significantly weakened when the special relativistic effects are considered. Compared with the case without special relativistic effects, when special relativistic effects are considered the winds are closer to the disc surface, the maximum speed of winds is reduced by $\sim$20 percent--70 percent, and the mass outflow rate and the kinetic power is significantly reduced.}
   {}
\keywords{accretion -- black hole physics -- hydrodynamics}
\maketitle

\section{Introduction} \label{sec:intro}

Observations indicate that winds or outflows are commonly present in various objects, including active galactic nuclei (AGNs), X-ray binaries, and early and late stars. In luminous AGNs, blueshifted absorption lines are often detected in their ultraviolet (UV) and X-ray spectra \citep{1999AJ....117.2573C,2004ApJ...601..715N,2010A&A...521A..57T,2011MNRAS.411..247R,2012MNRAS.422L...1T,2013MNRAS.430...60G}. For example, \cite{2007ApJS..171....1M} found from 37 optically bright quasars that 10\%--17\% of C IV narrow absorption lines in the UV band are blueshifted at 5000--70,000 km s$^{-1}$. \cite{2012MNRAS.422L...1T,2014MNRAS.443.2154T} reported that $>$35\% of radio-quiet AGNs have highly ionized Fe XXV and Fe XXVI absorption lines in the hard X-ray band, which blueshifted at $\sim$0.03--0.3 $c$ ($c$ is the speed of light). It was suggested that the Fe XXV and Fe XXVI absorption lines are generated by a highly ionized absorber from the accretion disc around the central black hole \citep[BH][]{2012MNRAS.422L...1T}. \cite{2010A&A...521A..57T} referred to the highly ionized absorbers with a velocity above 10$^4$ km s$^{-1}$ as UFOs. Based on photoionization modeling, \cite{2011ApJ...742...44T} derived the basic properties of ultra-fast outflows (UFOs). They found that the ionization parameter ($\xi$) and column density (N$_{\rm H}$) of UFOs are distributed in the range of log($\xi/(\text{erg}$ $\text{s}^{-1} \text{cm})$) $\sim$3-6 and log($N_{\rm{H}}/(\text{cm}^{-2})$) $\sim$22-24, respectively.

The formation of UFOs is still an open issue. Two kinds of models that produce UFOs from the accretion disc have been proposed. One is a line-force driving model, and the other is a magnetic-driving model \citep{2004ApJ...616..688P,2014ApJ...780..120F,2016PASJ...68...16N,2017MNRAS.465.2873N,2018ApJ...864L..27F,2021ApJ...914...31Y,2021MNRAS.503.1442M}. For the magnetic-driving model, it was proposed that when a poloidal component ($B_{\rm p}$) is weaker than the toroidal component ($B_{\phi}$), i.e. $|B_{\phi}/B_{\rm p}|\lesssim1$, the magneto-centrifugally driven mechanism drives winds \citep{1982MNRAS.199..883B}. When the toroidal magnetic field is much stronger than the poloidal component, i.e. $|B_{\phi}/B_{\rm p}|\gg1$, the magnetic pressure can drive winds. The magnetic-driving model has been developed
in previous works \citep[e.g.,][]{1987ApJ...315..504L,1987PASJ...39..821S,1994A&A...287...80C,2014ApJ...780..120F,2018ApJ...864L..27F}. \cite{2018ApJ...864L..27F} employed the magnetic-driving model to model the Fe K UFOs observed in PDS 456.

The line-force driving model has also been employed to understand the formation of UFOs \citep{1995ApJ...454L.105M,2000ApJ...543..686P,2016PASJ...68...16N,2021ApJ...914...31Y}. In luminous AGNs, a thin disc around a BH irradiates the gases above the disc, causing the photons from the disc surface to be scattered or absorbed by the gases, and then the Compton-scattering force and the spectral line force are exerted on the gas. For the weakly ionized gases, the line force caused by the absorption of UV photons becomes important and is the main force driving winds \citep{1975ApJ...195..157C}. For the highly ionized gas, the line force is negligible and the Compton-scattering force is the main radiation force. In the luminous AGNs, the Compton-scattering force is not strong enough to drive the winds with a mildly relativistic velocity, and then the line force is an important driving force because the thin disc can emit strong UV radiation \citep{1995ApJ...454L.105M}. When the ionization parameter of gases is higher than 100 erg s$^{-1}$ cm, the line force becomes unimportant. However, the ionization parameter of UFOs is much higher than 100 erg s$^{-1}$ cm. Numerical simulations found that the line force can intermittently accelerate winds and then form the observed UFOs \citep{2017MNRAS.465.2873N,2021MNRAS.503.1442M,2021ApJ...922..262Y,2021ApJ...914...31Y}. When the dense gases in the inner region of the disc shield the X-ray photons from the central corona, the gases are weakly ionized, and then the line force becomes effective and accelerates winds. When the winds driven by the line force are exposed to the X-ray photons, the winds are highly ionized and become UFOs. Therefore, UFOs may come of the line-force-driven winds\citep{2017MNRAS.465.2873N,2021MNRAS.503.1442M,2021ApJ...922..262Y,2021ApJ...914...31Y}. \cite{2014ApJ...789...19H} employed the post-processing approach to implement Monte Carlo simulations of radiation transfer. They found that the `shielding' effect is ineffective because scattered X-ray photons can contribute to ionizing the gas. In the work of \cite{2014ApJ...789...19H}, the simulation of radiation transfer and hydrodynamics was decoupled. The dynamical properties of winds would not change due to alterations in the radiation field. In reality, winds could adjust themselves to ensure that the scattered photons were not sufficiently strong, thereby maintaining the effectiveness of the shielding effect and allowing the line force to efficiently drive winds.

When the line-force-driven winds move at mildly relativistic velocities in the radiation field, special relativistic effects, such as the radiation-drag effect, stop being negligible. When winds move at relativistic velocities within a radiation field, a radiation-drag force acts on them due to relativistic effects. This effect is referred to as the radiation-drag effect. We also note another relativistic effect, namely the Poynting-Robertson effect. This effect arises when the radiation emitted by atoms in winds is collimated in the direction of motion of the atoms due to relativistic corrections, resulting in a drag force acting on the emitting atoms. \cite{2023A&A...670A.122M} have investigated the Poynting-Robertson effect on the black-hole-driven winds using an analytical approach. In this paper, we focus on the radiation-drag effect based on numerical simulations.

Previous works found that the wind speed is significantly reduced when the relativistic effects are taken into account \citep{2020A&A...633A..55L,2021A&A...646A.111L,2022MNRAS.515.5594W,2023MNRAS.518.2693Q}. The previous works employed radiation hydrodynamic simulations to study the line-force-driven winds in AGNs\citep[e.g.,][]{2000ApJ...543..686P,2004ApJ...616..688P,2017MNRAS.465.2873N,2021ApJ...922..262Y,2021ApJ...914...31Y}. However, the simulations ignored relativistic effects, such as the radiation-drag effect and the influence of the disc rotation on the radiation field. Therefore, our motivation in this paper is to study the influence of the relativistic effects on the line-force-driven winds.

The paper is organized as follows. In section 2, we describe basic equations and the model set-up. In sections 3 and 4, we give our results and application. In section 5, we give conclusions.

\section{Radiation field} \label{sec2.1}

A geometrically thin and optically thick disc is usually employed to describe luminous AGNs such as quasars \citep{1973A&A....24..337S}. Observations also indicate the presence of a hot corona within a Schwarzschild radius ($r_{\rm s}$) of 10 \citep{2013ApJ...769L...7R,2014A&ARv..22...72U}. We followed \citet{2000ApJ...543..686P} in assuming a model of a thin disc with a spherical hot corona to describe the luminous AGNs. The gases above the thin disc are irradiated by the radiation from the disc surface and the hot corona. On the thin disc, the local isotropic intensity, $I_{\rm d}(r_{\rm d})$, is written as
\begin{equation}
I_{\rm d}(r_{\rm d})=\frac{9\epsilon L_{\rm Edd}r_{\rm s}}{4 \pi r_{\rm d}^3}[1-(\frac{3r_{\rm s}}{r_{\rm d}})^{\frac{1}{2}}],
\end{equation}
where $r_{\rm d}$ is the radial position on the disc surface, $L_{\rm Edd}$ is the Eddington luminosity, and $\epsilon$ is the Eddington ratio of disc luminosity. The disc luminosity is given by $L_{\rm d}=\epsilon L_{\rm Edd}$. The disc can emit a large number of UV photons that contribute to the line force through the bound-bound transition. According to \cite{2000ApJ...543..686P} and \cite{2016PASJ...68...16N}, we roughly assume that the disc with a local temperature higher than $3\times10^3$ K emits all disc photons that can contribute to the line force. The hot corona emits the X-ray photons, which play an important role in ionizing gases. We define the corona luminosity as $L_{\rm cor}=f_{\rm x}L_{\rm d}$, where $f_{\rm x}$ is the ratio of the corona luminosity to the disc luminosity. In our models, the corona luminosity ($L_{\rm cor}$) is determined by the ratio ($f_{\text{x}}$) of the corona luminosity to the disc luminosity. We evaluated $f_{\rm x}$ according to an observed non-linear relation between the X-ray (2 keV) and UV (2500 {\AA}) emissions in quasars \citep{2016ApJ...819..154L}. This relation implies that in optically brighter AGNs the X-ray luminosity is fainter, i.e. $f_{\rm x}$ is less. According to this relation, \cite{2021ApJ...914...31Y} evaluated the $f_{\rm x}$. The calculation of $f_{\rm X}$ is referred to in \cite{2021ApJ...914...31Y}.

\begin{table}
    \begin{center}
    \caption{Basic parameters of simulation set-up}\label{tab1}
    \begin{tabular}{ccp{4.5cm}}
    \hline\noalign{\smallskip} \hline\noalign{\smallskip}
    Parameters & Value & Definition\\
    \hline\noalign{\smallskip}
    $M_{\rm bh}$ & $10^8 M_{\odot}$      & Mass of black hole   \\
    $\gamma$     & 5/3                & Adiabatic index      \\
    $\mu$          & 1.0               & Mean molecular   \\
    $T_{\rm x}$    & 10$^8$ K            & Characteristic temperature of X-ray radiation \\
    \hline\noalign{\smallskip}
    \end{tabular}
    \end{center}
\end{table}

The line-force-driven winds from the disc have mildly relativistic velocities and the special relativistic effects cannot be ignored. When we calculate the radiation flux from the thin disc, the transformation of the frame needs to be considered. There are two different frames here: one is the co-moving frame moving with gases and the other is the rest frame standing at infinity. In the rest frame, the radiation intensity from the disc surface is given by
\begin{equation}
I_0=\frac{I_{\rm d}}{(1-\frac{\overrightarrow{\textit{v}_{\rm d}}\cdot\overrightarrow{\textit{l}}}{c})^4}, \label{eq2}
\end{equation}
where $\overrightarrow{\textit{l}}$ is the direction-cosine vector pointing toward the rest observer from the disc surface and $\overrightarrow{\textit{v}_{\rm d}}$ is the disc velocity. The denominator on the right side of equation (\ref{eq2}) reflects the special relativistic effect caused by disc rotation. The radiation energy density ($E_{0}$), the radiative flux ($F_{i,0}$), and the radiation stress tensor ($P_{ij,0}$) in the rest frame read
\begin{equation}
E_0=\frac{1}{c}\oint_{\Omega} I_0  d\Omega,
\end{equation}
\begin{equation}
F_{i,0}=\oint_{\Omega} I_0 l_{i} d\Omega,
\end{equation}
and
\begin{equation}
P_{ij,0}=\frac{1}{c}\oint_{\Omega} I_0 l_{i}l_{j} d\Omega,\label{eq5}
\end{equation}
where $c$ is the speed of light and $\Omega$ is the solid angle of the disc toward the rest observer. When we calculate the radiation force exerted on the gases, we need to transform the radiation flux generated by the disc from the rest frame into the co-moving frame. On the order of $v/c$, the transformation of the two frames is given by
\begin{equation}
F_{i}=F_{i,0}-E_{0}v_{i}- \sum_{j}v_{j}P_{ij,0},
\label{transform}
\end{equation}
where $F_{i}$ is the radiative flux in the co-moving frame and $v_{i}$ is the $i$th component of the gas velocities. In the above equations, the subscript `0' denotes the quantities in the rest frame. In equation (\ref{transform}), the $(-E_{0}v_{i}- \sum_{j}v_{j}P_{ij,0})$ term reflects the special relativistic effect caused by the motion of the gas, and means that the radiation flux measured by the gas moving in the direction of the radiation flux decreases and then the radiation force exerted on the moving gas weakens. This effect is called the radiation-drag effect.

\subsection{Basic equations} \label{sec2.2}

\begin{table*}[htbp]

     \centering
    \caption{Summary of models}\label{tab2}
    \hspace*{-0.8cm}
    \begin{tabular}{c c c c c c c c}
    \hline
    models & $\epsilon$ & $\dot{M}_{\rm w}$ & $P_{\rm m}$ & $P_{\rm k}$  & $P_{\rm th}$ & $\theta_{\rm out}$ & $v_{\rm max}$ \\
           & ($L_\text{Edd}$) & ($L_\text{Edd}/c^2$) & ($L_\text{Edd}/c$) & ($L_\text{Edd}$) & ($L_\text{Edd}$) & & ($c$) \\
     (1)& (2) & (3) & (4) & (5) & (6) & (7) & (8) \\
    \hline
     A1 & 0.1 & $3.90\times10^{-2}$ & $4.53\times10^{-4}$& $2.66\times10^{-5}$  & $1.94\times10^{-7}$ &14.9$^{\circ}$-24.0$^{\circ}$ & 0.046 \\
     A2 & 0.2 & $4.95\times10^{-1}$ & $1.67\times10^{-2}$& $2.86\times10^{-3}$  & $4.19\times10^{-6}$ &21.2$^{\circ}$-50.6$^{\circ}$ & 0.114 \\
     A3 & 0.3 & $6.27\times10^{-1}$ & $2.86\times10^{-2}$& $6.54\times10^{-3}$  & $5.75\times10^{-6}$ &24.0$^{\circ}$-55.5$^{\circ}$ & 0.188 \\
     A4 & 0.4 & $6.16\times10^{-1}$ & $3.08\times10^{-2}$& $7.90\times10^{-3}$  & $6.59\times10^{-6}$ &25.6$^{\circ}$-55.5$^{\circ}$ & 0.186 \\
     A5 & 0.5 & $9.21\times10^{-1}$ & $4.30\times10^{-2}$& $1.06\times10^{-2}$  & $1.08\times10^{-5}$ &24.8$^{\circ}$-66.6$^{\circ}$ & 0.176 \\
     A6 & 0.6 & $1.58$ & $1.28\times10^{-1}$& $5.47\times10^{-2}$  & $1.52\times10^{-6}$ &28.1$^{\circ}$-42.1$^{\circ}$ & 0.484 \\
     A7 & 0.7 & $1.83$ & $1.73\times10^{-1}$& $8.18\times10^{-2}$  & $7.95\times10^{-7}$ &28.1$^{\circ}$-43.4$^{\circ}$ & 0.438 \\
     A8 & 0.8 & $1.95$ & $2.75\times10^{-1}$& $2.08\times10^{-1}$  & $9.76\times10^{-5}$ &28.1$^{\circ}$-57.3$^{\circ}$ & 0.563 \\ \\
     \hline
    B1 & 0.1 &  / & / &  / & / & / & / \\
    B2 & 0.2 & $4.76\times10^{-2}$ & $6.34\times10^{-4}$ & $4.46\times10^{-5}$  & $2.22\times10^{-7}$ &10.1$^{\circ}$-27.2$^{\circ}$ & 0.042 \\
    B3 & 0.3 & $2.37\times10^{-1}$ & $6.67\times10^{-3}$ & $9.55\times10^{-4}$  & $1.31\times10^{-6}$ &16.5$^{\circ}$-36.0$^{\circ}$ & 0.107 \\
    B4 & 0.4 & $3.55\times10^{-1}$ & $1.38\times10^{-2}$ & $2.74\times10^{-3}$  & $1.71\times10^{-6}$ &15.4$^{\circ}$-43.4$^{\circ}$ & 0.146 \\
    B5 & 0.5 & $8.54\times10^{-1}$ & $3.07\times10^{-2}$ & $5.57\times10^{-3}$  & $9.09\times10^{-6}$ &18.7$^{\circ}$-32.8$^{\circ}$ & 0.146 \\
    B6 & 0.6 & $8.98\times10^{-1}$ & $5.76\times10^{-2}$ & $1.90\times10^{-2}$  & $1.80\times10^{-6}$ &24.8$^{\circ}$-31.8$^{\circ}$ & 0.341 \\
    B7 & 0.7 & $1.23$ & $5.88\times10^{-2}$ & $1.50\times10^{-2}$  & $4.71\times10^{-6}$ &17.0$^{\circ}$-28.1$^{\circ}$ & 0.334 \\
    B8 & 0.8 & $1.30$ & $8.70\times10^{-2}$ & $2.92\times10^{-2}$  & $2.64\times10^{-7}$ &18.1$^{\circ}$-30.9$^{\circ}$ & 0.401 \\ \\
    \hline
    C1 & 0.1 & / & / & / & / & / & / \\
     C2 & 0.2 & $5.37\times10^{-2}$ & $7.89\times10^{-4}$& $5.83\times10^{-5}$  & $1.21\times10^{-7}$ &12.3$^{\circ}$-20.5$^{\circ}$ & 0.051 \\
     C3 & 0.3 & $1.79\times10^{-1}$ & $3.37\times10^{-3}$& $3.20\times10^{-4}$  & $6.95\times10^{-7}$ &13.1$^{\circ}$-34.9$^{\circ}$ & 0.073 \\
     C4 & 0.4 & $3.40\times10^{-1}$ & $1.08\times10^{-2}$& $1.75\times10^{-3}$  & $1.38\times10^{-6}$ &17.0$^{\circ}$-42.1$^{\circ}$ & 0.129 \\
     C5 & 0.5 & $4.41\times10^{-1}$ & $1.68\times10^{-2}$& $3.14\times10^{-3}$  & $1.49\times10^{-6}$ &15.9$^{\circ}$-43.4$^{\circ}$ & 0.136 \\
     C6 & 0.6 & $4.45\times10^{-1}$ & $1.75\times10^{-2}$& $3.49\times10^{-3}$  & $2.39\times10^{-6}$ &18.1$^{\circ}$-46.2$^{\circ}$ & 0.142 \\
     C7 & 0.7 & $5.60\times10^{-1}$ & $1.21\times10^{-2}$& $1.67\times10^{-3}$  & $2.41\times10^{-6}$ &14.0$^{\circ}$-26.4$^{\circ}$ & 0.131 \\
     C8 & 0.8 & $1.26$ & $3.31\times10^{-2}$& $6.17\times10^{-3}$  & $3.54\times10^{-6}$ &17.5$^{\circ}$-24.0$^{\circ}$ & 0.210 \\ \\
    \hline
    \end{tabular}
    \begin{tablenotes}
    \item Notes: Column (1): run names; Column (2): the ratio of disc luminosity to Eddington luminosity; Column (3): the time-averaged values of the mass outflow rate; Column (4)-(6): the time-averaged values of the momentum flux, the kinetic flux, and thermal energy flux, which are carried out by winds at the outer boundary; Column (7): the azimuth range of winds at the outer boundary; Column (8): the maximum speed of winds. Runs A1--A8 ignore the special relativity effects caused by disc rotation and gas motion. The A-group runs is identical to the models proposed by \cite{2000ApJ...543..686P}. Runs B1--B8 take into account the special relativity effect caused by disc rotation but ignore the radiation-drag term in equation (\ref{transform}). Runs C1--C8 take into account the special relativity effect caused by disc rotation and the radiation-drag effect.
    \end{tablenotes}
\end{table*}

To simulate the irradiated gases above the disc surface, we placed the computation region above the disc surface and implemented our simulations in spherical polar co-ordinates ($r$,$\theta$,$\phi$). We placed the $\theta=\pi/2$ plane at the disc surface and assumed that the disc scale height is constant with radius. Following \cite{2016PASJ...68...16N},  we set the scale height of the thin disc to be $H_0=3.1 \epsilon r_{\rm s}$. We axis-symmetrically solved the hydrodynamical (HD) equations as follows:

\begin{equation}
    \frac{d \rho}{d t} +\rho \nabla \cdot  \overrightarrow{v} = 0 ,
    \label{eq7}
\end{equation}
\begin{equation}
    \rho \frac{d\overrightarrow{v}}{dt} = -\nabla P + \rho \nabla\psi +\rho \bf{g}^{rad},
     \label{eq8}
\end{equation}
\begin{equation}
    \rho \frac{d}{dt} (\frac{e}{\rho}) = -P \nabla \cdot \overrightarrow{v} + \rho \mathscr{L},
    \label{eq9}
\end{equation}
where $d/dt$($\equiv \partial/\partial+\overrightarrow{v}\cdot\bf{\bigtriangledown}$) denotes the Lagrangian derivative. The $\rho$, $P$, $\overrightarrow{v}$, and $e$ in equations \ref{eq7}--\ref{eq9} are density, pressure, velocity, and internal energy, respectively. An equation of state of ideal gases, $P=(\gamma-1)/e $, was applied and $\gamma=5/3$ was set. $\psi$ is the BH pseudo-Newtonian potential \citep{1980A&A....88...23P} and then $\psi=-GM_{\rm BH}/(r_{\rm c} -r_{\rm s})$, where $G$ and $M_{\rm BH}$ are the gravitational constant and the BH mass, respectively, and $r_{\rm c}=\sqrt{r^2+H_{0}^2-2rH_{0}cos(\pi-\theta)}$ is the distance from the co-ordinate origin (the centre of the BH) to the gases. In this paper, we assume $M_{\rm BH} =10^8M_{\odot}$.

The $\mathscr{L}$ in equation (\ref{eq9}) is the net cooling rate per unit mass, which is referred to equations (18)-(21) in \cite{2000ApJ...543..686P}. The equations given by \cite{2000ApJ...543..686P} include Compton heating and cooling, X-ray ionization and recombination cooling, Bremsstrahlung cooling, and line cooling. The net cooling rate depends on the gas temperature ($T$), the ionization parameter ($\xi$), and the characteristic temperature ($T_{\rm x}$) of X-ray radiation. The gas temperature is given by $T=\mu m_{\rm p} P/k_{\rm B} \rho$, where $m_{\rm p}$, $k_{\rm B}$, and $\mu$ are the proton mass, the Boltzmann constant, the mean molecular weight, respectively. The ionization parameter, $\xi$, is defined as $\xi=e^{-\tau_{\text{x}}}L_{\text{cor}}/(r^2 n)$, where $\tau_{\rm x}$ is the optical depth of the X-ray and $n$ is number density ($n=\rho/\mu m_{\text{p}}$) of the gases. We followed \cite{2000ApJ...543..686P} in setting $\mu=1$ and $T_{\rm x}=10^8$ K.

The $\bf{g}^{\rm rad}$ in equation (\ref{eq8}) is the acceleration produced by both the line force and the Compton-scattering force. The $i$th component of radiation force ($\bf{g}^{\rm rad}$ ) was evaluated with
\begin{equation}
    {g}^{\rm{rad}}_{i} = e^{-\tau_{\rm d}}{\overset{}{\oint_{\Omega}}} \frac{\sigma_{\text{e}}}{\textit{c}}d{F}_{\textit{i}}+
    e^{-\tau_{\rm d}}{\overset{}{\oint_{\Omega}}} \mathcal{M}\frac{\sigma_{\textit{e}}}{\textit{c}}d{F}_{\textit{i}}^{\rm{UV}}
\label{radforce}
,\end{equation}
where the first term on the right side is the Compton-scattering force (${g}_{\textit{i}}^{\rm rad,e}$) produced by the all the radiation from the disc surface and the second term is the line force (${g}_{\textit{i}}^{\rm{rad},\textit{l}}$ ) produced by the all the radiation from the hot region ($\geq 3000$ K) of the disc surface. In the equation (\ref{radforce}), $\tau_{\rm d}$ is the optical depth of the radiation from the disc, $\sigma_{\rm e}$ is the mass-scattering coefficient of electrons, and $\mathcal{M}$ is the force multiplier of line force. $d{F}_{i}$ and $dF_{i}^{\rm{UV}}$ are the fluxes measured in the comoving frame with gases. According to equation (\ref{transform}), the Compton-scattering force is written as
\begin{equation}
{g}_{\textit{i}}^{\rm rad,e}= e^{-\tau_{\rm d}}\frac{\sigma_{\text{e}}}{c}(F_{i,0}-E_{0}v_{i}- \sum_{j}v_{j}P_{ij,0}).
\label{scforce}
\end{equation}
The details of the calculations about ${g}_{\textit{i}}^{\rm{rad},\textit{l}}$  are referred to in Appendix \ref{appendix}. To evaluate $\tau_{\rm x}$ and $\tau_{{\rm d}}$, we set the attenuation of the disc radiation to be $0.4 \text{ g}^{-1}\text{cm}^2$ and the X-ray attenuation to be $0.4 \text{ g}^{-1}\text{cm}^2$ for $\xi\geq10^5$ erg s$^{-1}$ cm but $40 \text{ g}^{-1}\text{cm}^2$ for $\xi<10^5$ erg s$^{-1}$ cm.

The numerical procedure of solving equations (\ref{eq7})--(\ref{eq9}) was divided into the following steps: (1) solving HD terms, (2) implicitly updating the gas temperature according to the net cooling rate, (3) explicitly updating the gas velocity according to the gravity and the radiation force. In the first step, a computational HD code, called Virginia Hydrodynamics One (VH1), was employed to evaluate HD terms. The VH1 code employs the piecewise parabolic method \citep{1984JCoPh..54..174C}.

\subsection{Model set-up} \label{sec2.3}

The computational domain covers the radial range of $r_{\rm in} (=30r_{\rm s}) \le r \le r_{\rm out}(=1500r_{\rm s})$ and the angular range of $0\le\theta\le \pi/2$. We divided the computational domain into 144$\times$160 zones. In the $r$ direction, we used 144 zones that were non-uniformly distributed in the range of $r_{\rm in}\le r \le r_{\rm out}$ and set the radial size ratio to be $(\Delta r)_{i+1}/(\Delta r)_{i} = 1.04$. In the $\theta$ direction, we used the 16 uniform zones in an angular range of $0^\circ$-$15^\circ$ and the 144 non-uniform zones in an angular range of $15^\circ$-$90^\circ$ and set the angular size ratio to be $(\Delta \theta)_{j+1}/(\Delta \theta)_{j} = 0.970072$. The minimum angular size is then $\Delta \theta= 0.0117947^\circ$ at the plane of $\theta =  \pi/2$.

Initially, we followed \cite{1998MNRAS.295..595P} in assuming that the gases are in a state of hydrostatic equilibrium in the vertical direction. The initial density distribution is then given by
\begin{equation}
    \rho(r,\theta) = \rho_{\rm d} \text{exp} (-\frac{GM_{\rm bh}}{2 c_{\rm s,d}^2 r (1-\frac{r_{\rm s}}{r})^2\text{tan}^2(\theta)}) \label{(10)},
\end{equation}
where $c_{\rm s, d}$ and $\rho_{\rm d}$ are the isothermal sound speed and the density at the disc surface, respectively. We used the local effective temperature ($T_{\rm eff}(r_{\rm d}) = (\pi I_{\rm d}/\sigma)^{1/4}$) at the disc surface to determine the isothermal sound speed. The initial temperature ($T(r,\theta)$) of gases was simply set to be $T(r,\theta)=T_{\rm eff}(r_{\rm d})$. The initial radial and polar velocities were set to zero. The rotational velocity is given by $v_\phi(r,\theta)= \sqrt{GM_{\rm bh}/r} \text{sin}\theta r /(r-r_{\rm s})$ according to force balance.

For the boundary conditions, we applied the outflow boundary condition at the inner and outer radial boundaries and employed the axially symmetric boundary condition at the pole. At the plane of $\theta=\pi/2$, we fixed density and temperature during the simulations. We set the density to be $\rho_{\rm d}=10^{-12}$ g cm$^{-3}$ and the temperature to be $T_{\rm eff}(r_{\rm d})$.

\section{Result} \label{sec3}

\begin{figure}[ht]
\includegraphics[width=.47\textwidth]{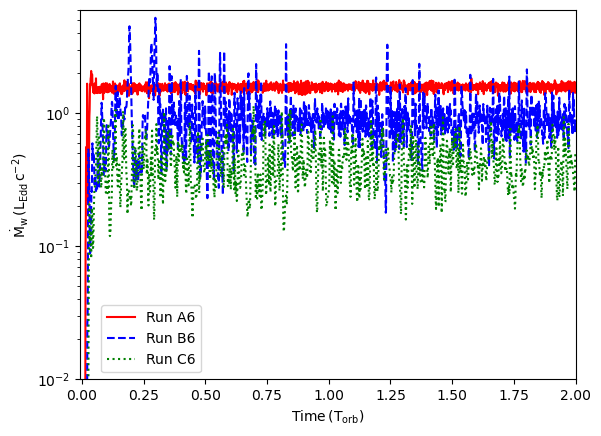}
	\ \centering \caption{ Time evolution of the mass outflow rate ($\dot{M}_{\rm w}$) at the outer boundary for runs A6, B6, and C6. The time unit ($T_{\rm orb}$) is the Keplerian orbital period at the outer boundary. }\label{fig.1}
\end{figure}

Table \ref{tab1} lists the basic parameters of models and Table \ref{tab2} summarizes our runs. In Table \ref{tab2}, columns (1) and (2) give run names and the disc luminosity in units of Eddington luminosity, respectively, columns (3)-(6) give the time-average quantities of winds at the outer boundary, such as the mass outflow rate ($\dot{M}_{\rm w}$), the momentum flux ($P_{\rm m}$), the kinetic energy flux ($P_{\rm k}$), and the thermal energy flux $P_{\rm th}$, and columns (7)-(8) give the azimuth range of winds and the maximum velocity, respectively. We divided our runs into three groups, i.e. A1--A8, B1--B8, and C1--C8. The A-group runs (runs A1-A8) ignore the special relativity effects caused by disc rotation and gas motion. When we set the denominator on the right of equation (\ref{eq2}) to be 1, we ignored the special relativity effect caused by disc rotation. When we ignored the radiation-drag term ($-E_{0}v_{i}- \sum_{j}v_{j}P_{ij,0}$) in equation (\ref{transform}), we ignored the special relativity effects caused by gas motion. Therefore, the A-group runs are identical to the models proposed by \cite{2000ApJ...543..686P}. The B-group runs (runs B1-B8) take into account the special relativity effects caused by disc rotation, while ignoring the radiation-drag term in equation (\ref{transform}). The C-group runs (runs C1-C8) take into account all special relativity effects, i.e. the special relativity effects caused by disc rotation and the radiation-drag effect.

\subsection{Fiducial runs}

Figure \ref{fig.1} shows the time evolution of the mass outflow rate ($\dot{M}_{\rm w}$) at the outer boundary for runs A6, B6, and C6. From figure \ref{fig.1}, a quasi-steady state can quickly be reached. Therefore, time averaging is implemented over the time interval of 1.0-1.5 $T_{\rm orb}$, where $T_{\rm orb}$ is the Keplerian orbital period at the outer boundary.

\begin{figure*}[bp]
\centering
\includegraphics[width=.3\textwidth]{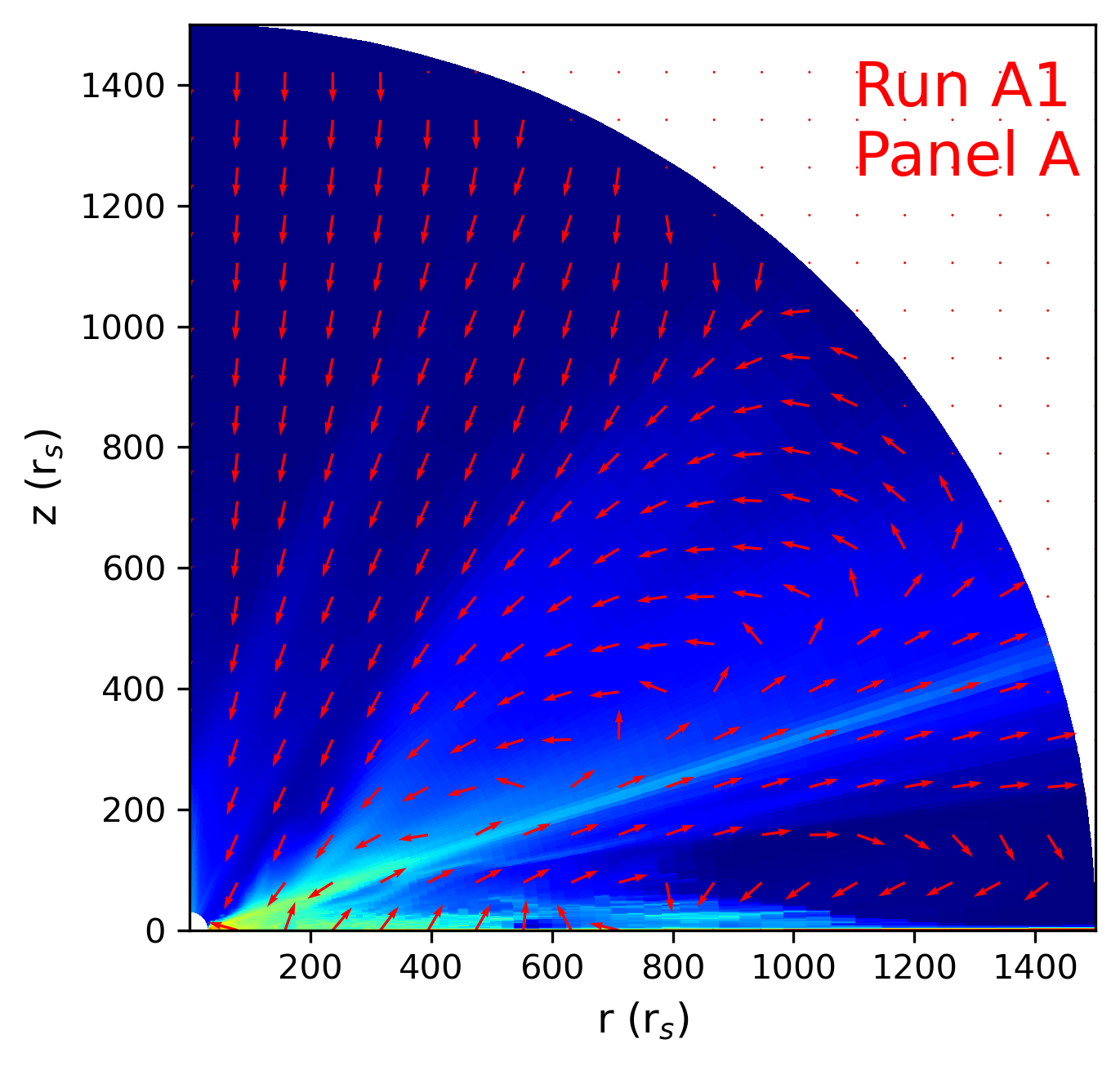}
\includegraphics[width=.3\textwidth]{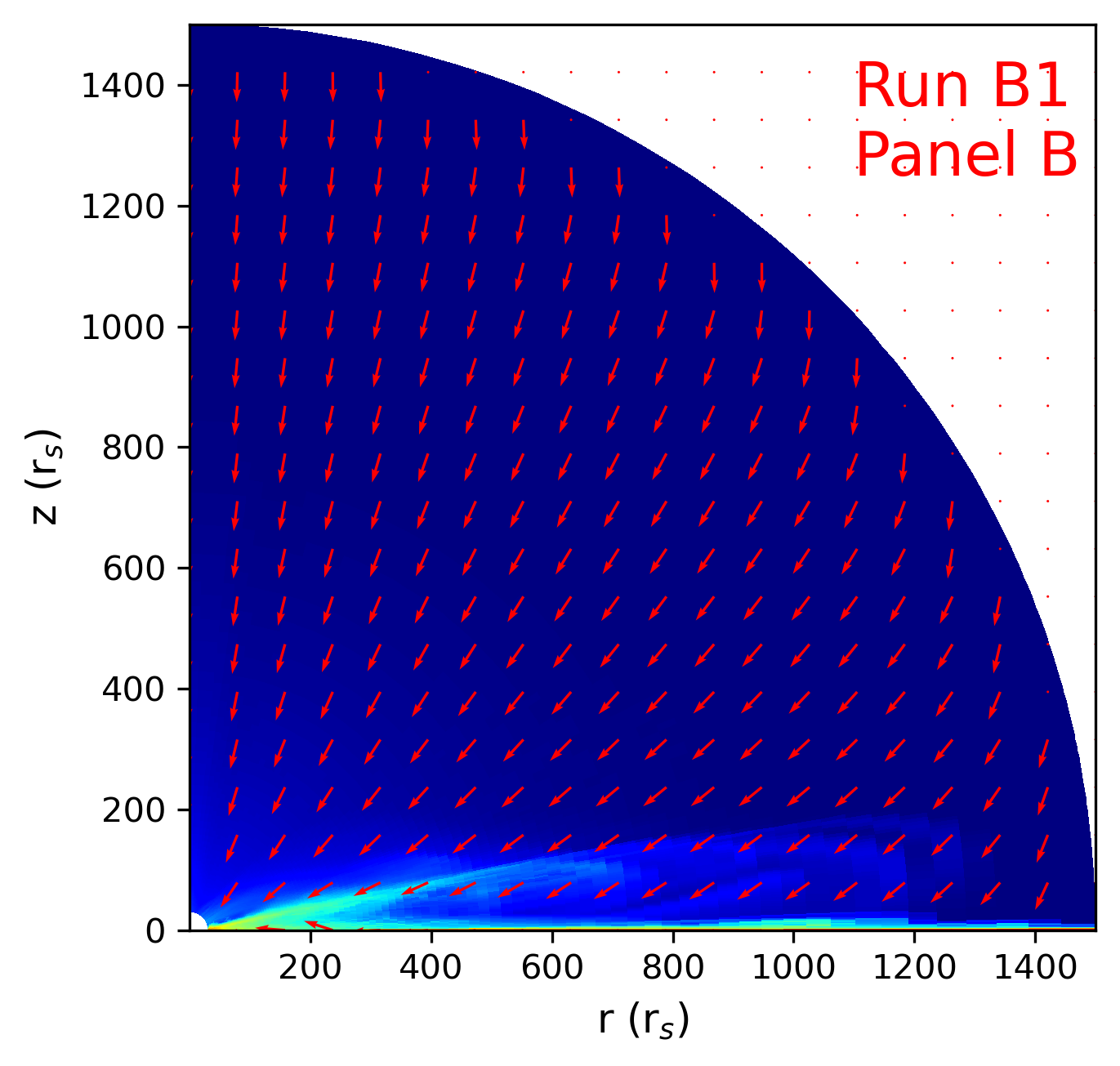}
\includegraphics[width=.375\textwidth]{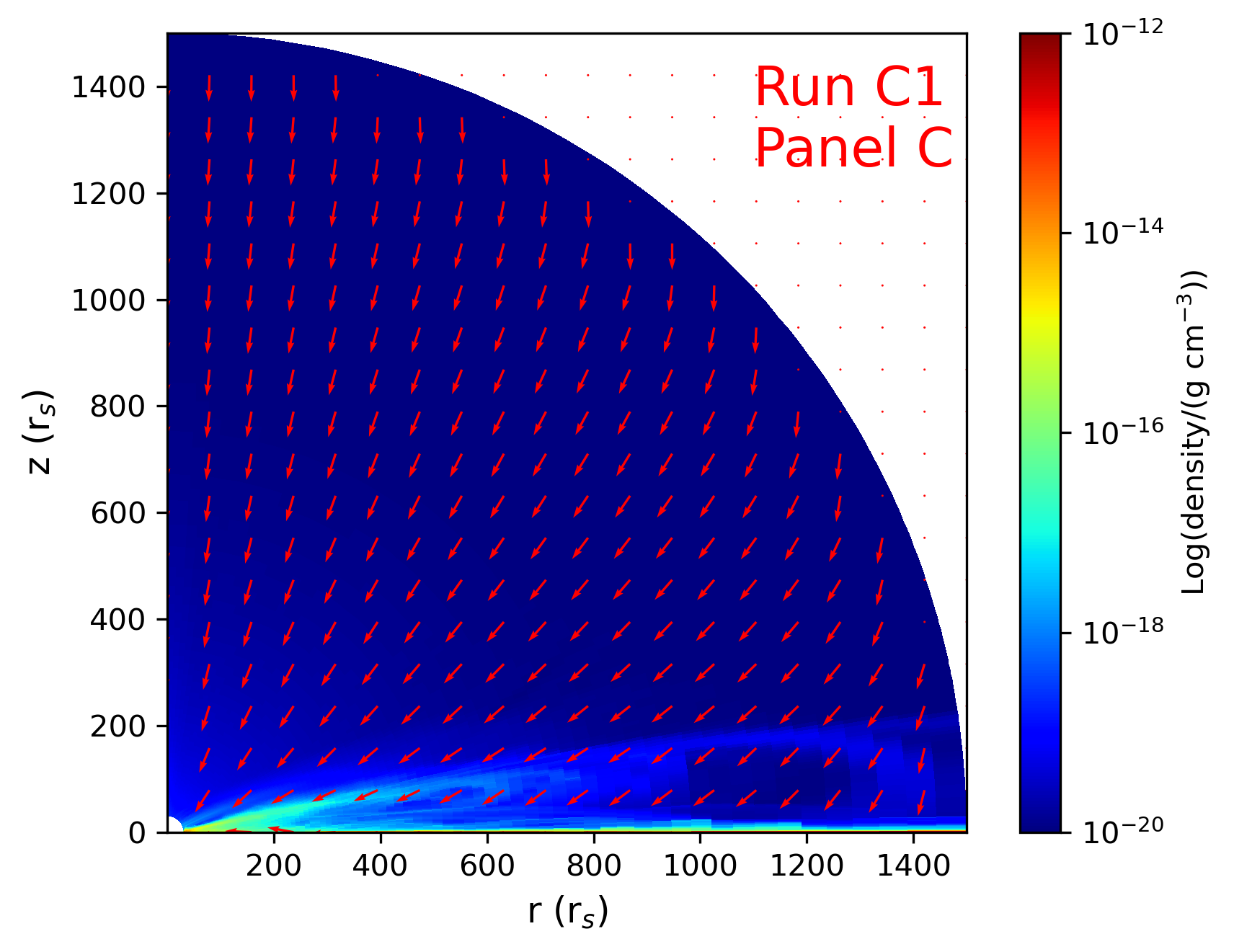}

\includegraphics[width=.3\textwidth]{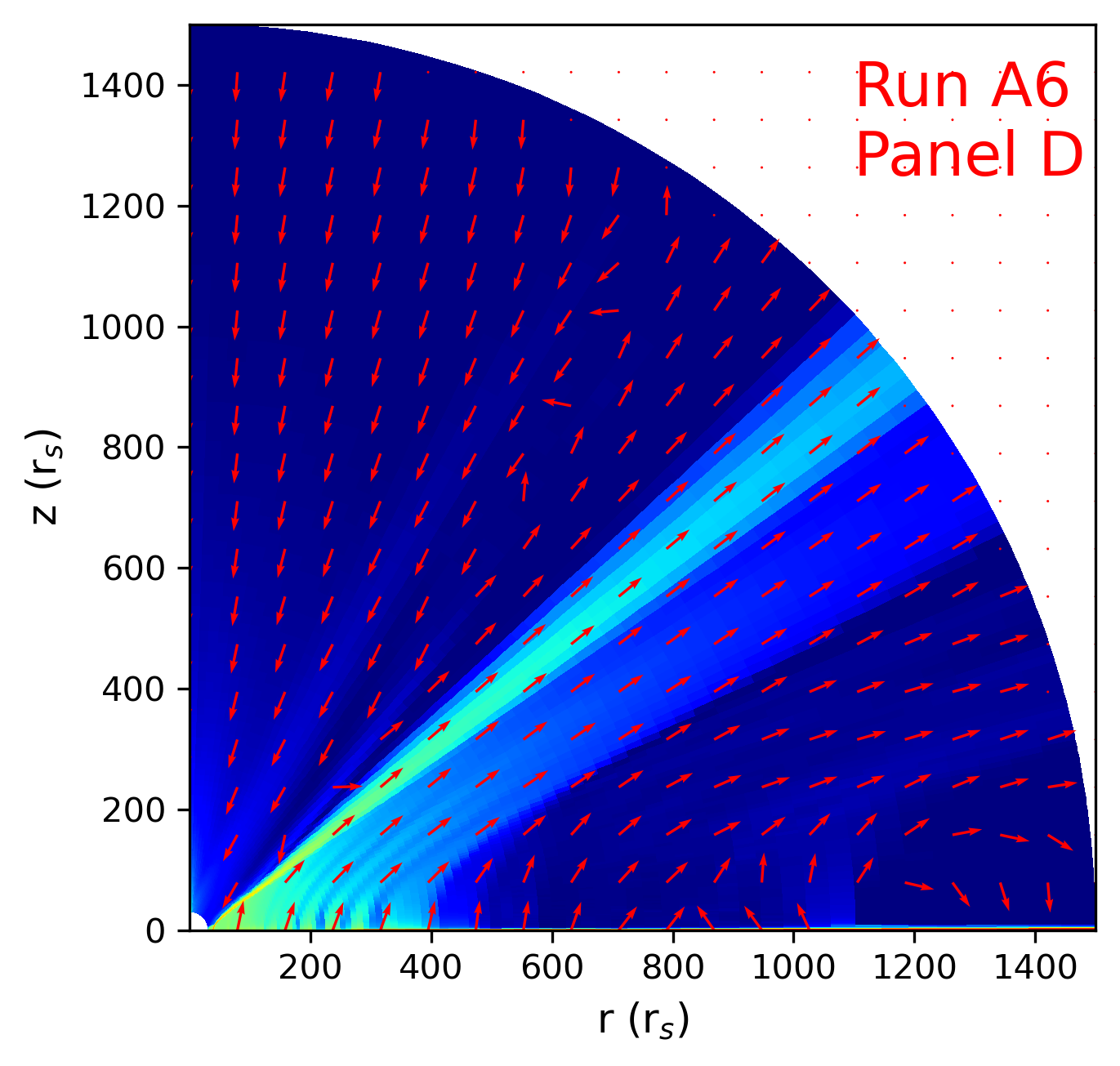}
\includegraphics[width=.3\textwidth]{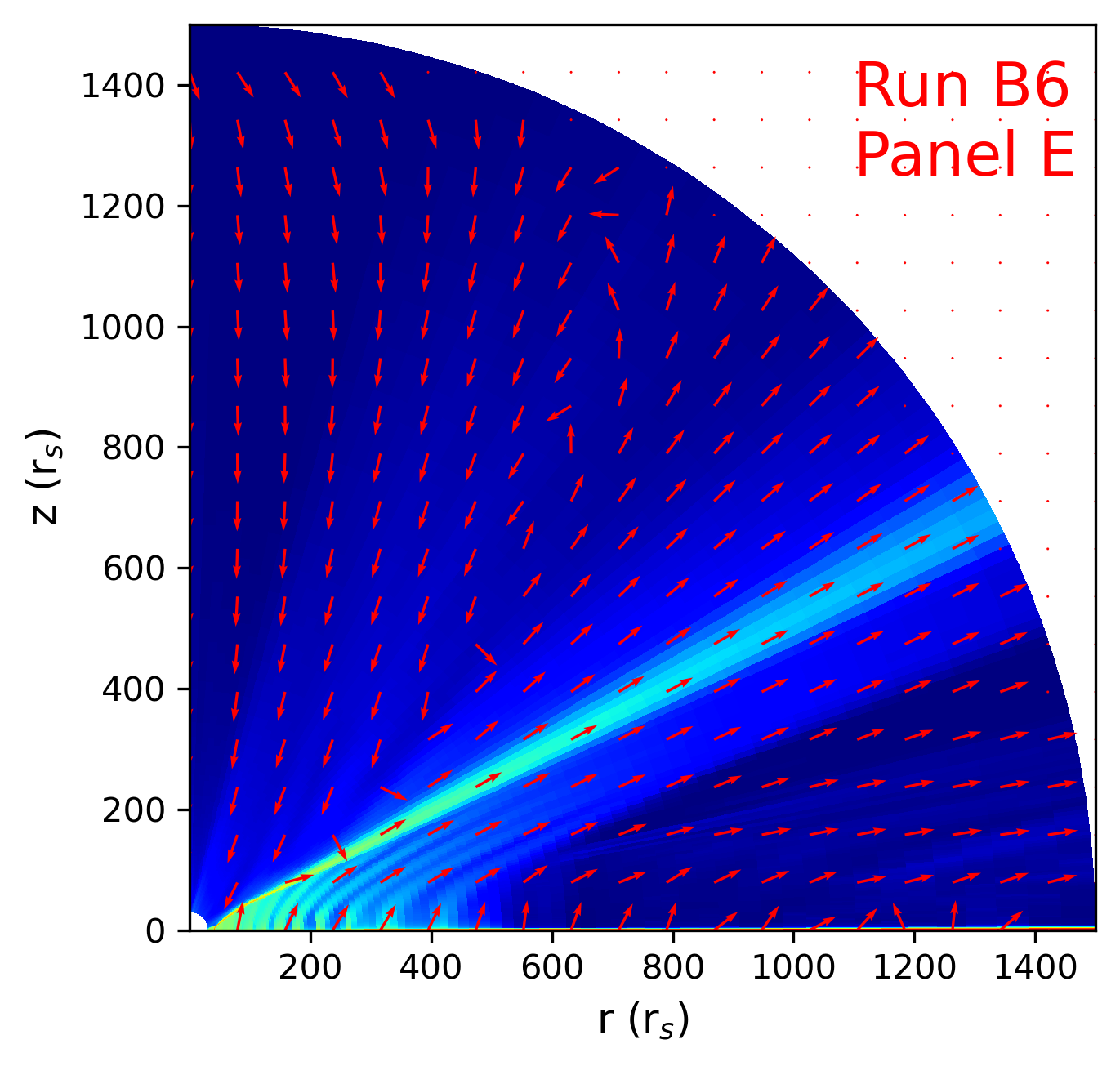}
\includegraphics[width=.375\textwidth]{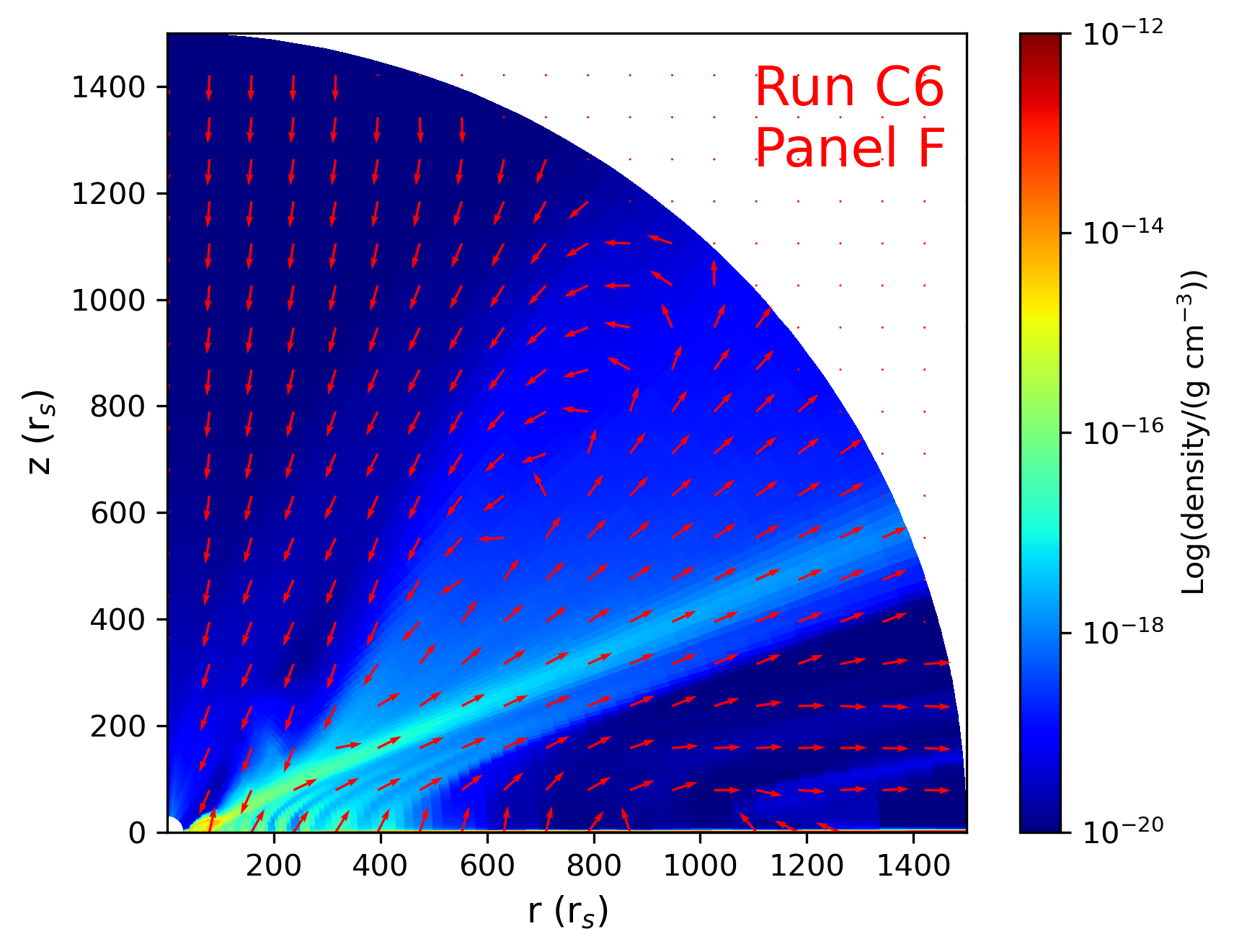}
    \caption{Two-dimensional structure of winds for runs A1, B1, C1, A6, B6, and C6. In panels A--F, colors mean the density distribution and arrows mean the poloidal velocity. The structure of winds is obtained by averaging over the time interval of 1.0--1.5 $T_{\rm orb}$. }\label{fig.2}
\end{figure*}

Figure \ref{fig.2} shows the poloidal distribution of wind density for runs A1, B1, C1, A6, B6, and C6. As is shown figure \ref{fig.2}, when the disc luminosity is equal to 0.1$L_{\rm Edd}$, runs B1 and C1 cannot generate winds that reach the outer boundary. In comparison to the wind observed in run A1, the failed wind is closer to the surface of disc. When the disc luminosity is higher then 0.2$L_{\rm Edd}$, all the runs generate high-velocity winds that can run away out the outer boundary. Observations have shown that UFOs are also detected in low-Eddington AGNs, such as NGC 2992, which has a luminosity of 0.02 $L_{\rm Edd}$ \citep{2018MNRAS.478.5638M}. Our results imply that for objects with a luminosity lower than 0.1$L_{\rm Edd}$, the line force plays little role in launching powerful outflows and that magnetic driving is a potential mechanism.

From figure \ref{fig.2}, we can see that the winds in runs B6 and C6 are closer to the disc surface than the winds in run A6. To understand this result, we drew figure \ref{fig.2.1}, which shows the ratio of the radiation fluxes and the line-force multiplier. Panels (A) and (B) in figure \ref{fig.2.1} imply that when the special relativity effects (i.e. the Doppler effect) caused by disc rotation are taken into account, the radiation flux strengthens. However, as is shown in panel (C), when the special relativity effects are taken into account, the line-force multiplier is reduced over the range of $\sim83^{\circ}$--90$^{\circ}$. As a result, the gases in runs B6 and C6 are not lifted up to a higher latitude. In comparing runs B6 and C6, the line-force multiplier for run C6 is significantly weaker than that for run B6 over the range of $\sim87^{\circ}$ to $90^{\circ}$. This results in the gases in run C6 being lifted up to a lower latitude than the ones in run B6.

\begin{figure*}[bp]
\centering
\includegraphics[width=.3\textwidth]{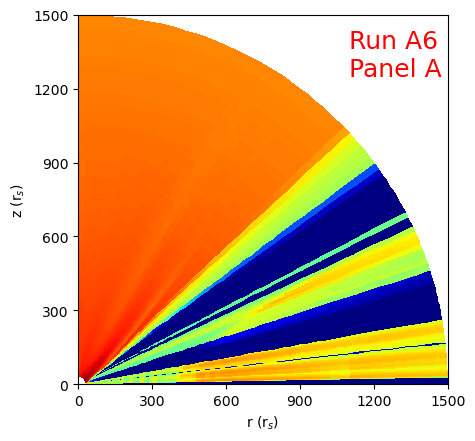}
\includegraphics[width=.3\textwidth]{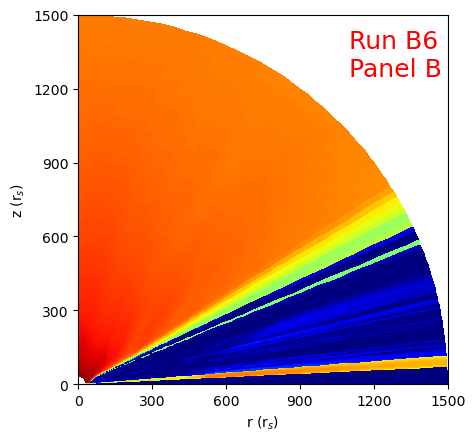}
\includegraphics[width=.35\textwidth]{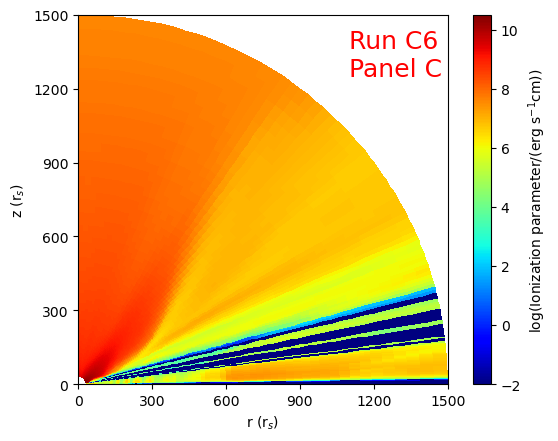}
    \caption{Two-dimensional distribution of the ionization parameter for runs A6, B6, and C6. The two-dimensional distribution is obtained by averaging over the time interval of 1.0--1.5 $T_{\rm orb}$. }\label{fig.2.0}
\end{figure*}

\begin{figure*}[ht]
\includegraphics[width=.3\textwidth]{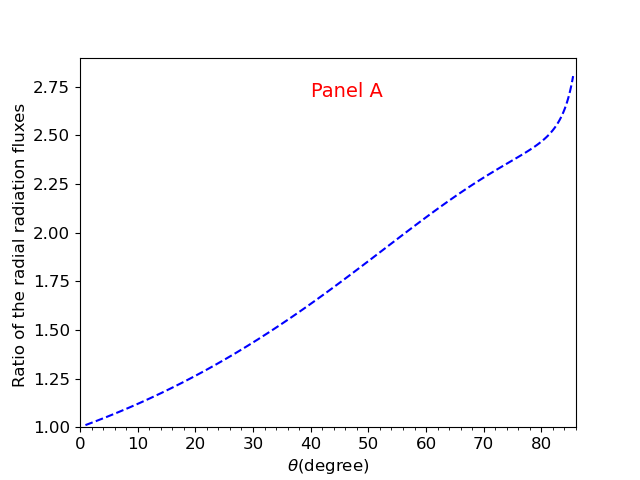}
\includegraphics[width=.3\textwidth]{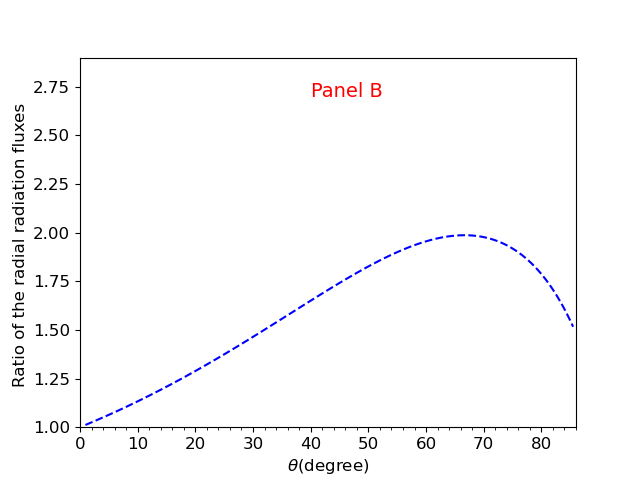}
\includegraphics[width=0.28\textwidth]{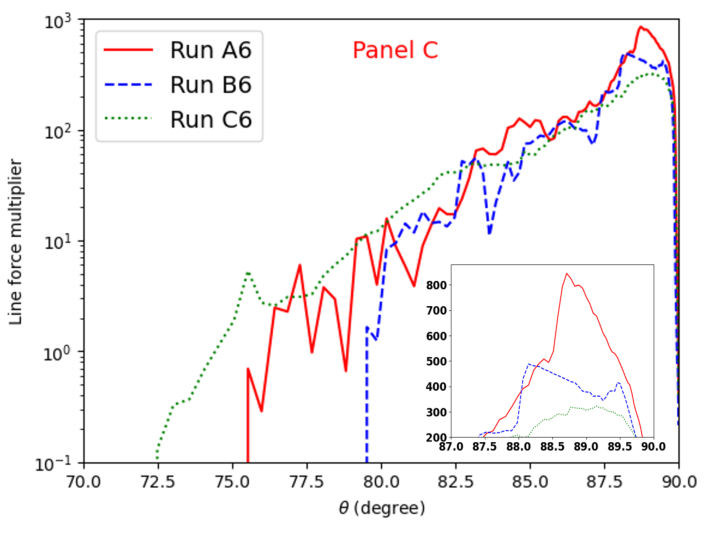}
	\ \centering \caption{The ratio of the radiation fluxes and the line-force multiplier. Panel (A) shows the ratio of the radial radiation fluxes at the outer boundary. Panel (B) shows the ratio of the angular radiation fluxes at $r=80 r_{\rm s}$. Panel (C) shows the angular distribution of the line-force multiplier of runs A6, B6, and C6 at $r=80 r_{\rm s}$. The ratio in panels (A) and (B) is calculated by dividing the radiation flux that incorporates the Doppler effect due to disc rotation by the radiation flux that excludes the Doppler effect.}
 \label{fig.2.1}
\end{figure*}

Figure \ref{fig.2.0} shows the poloidal distribution of the ionization parameter of winds for runs A6, B6, and C6. As is shown in figure \ref{fig.2.0}, the winds are extremely ionized in the polar region due to a very low column density. In the mid- to low-latitude regions, the ionization parameter undergoes a sudden change, which implies that a corresponding abrupt variation also occurs in X-ray radiation. The reasons are as follows. When the X-ray photons emitted from the central corona are shielded by the dense gases in the inner region at a specific angular position, the gases at the corresponding angular position exhibit low ionization. When the gases in the inner region are not dense enough to effectively shield the X-ray photons, the gases in the outer region are effectively ionized. Due to the scattered X-ray photons being neglected in our simulations, the high- and low-ionization regions exhibit a filament structure in the mid- to low-latitude regions. The distribution of the dense gases in the inner region evolves over time, which causes the angular position of filaments to change with time. After time-averaging, the ionization parameter also exhibits discontinuous variations in the mid- to low-latitude regions.

For runs A6, B6, and C6, we calculated the radial dependence of the mass outflow rate ($\dot{M}_{\rm w}(r)$), the momentum flux ($P_{\rm m}(r)$) of winds, and the kinetic power ($P_{\rm k}(r)$) of winds. According to Table \ref{tab2}, the power of winds is mainly determined by the kinetic power and then the thermal energy flux can be ignored. Here, we investigate the radial dependence of the thermal energy of winds. $\dot{M}_{\rm w}(r)$, $P_{\rm m}(r)$, and $P_{\rm k}(r)$ are given by
\begin{equation}
    \dot{M}_{\rm w}(r)=4\pi r^2 \int_{0^\circ}^{89.9^\circ} \rho \text{ max}(v_r,0) \text{ sin}\theta \text{ }d\theta ,
\end{equation}
\begin{equation}
    P_{\rm m}(r)=4\pi r^2 \int_{0^\circ}^{89.9^\circ} \rho \text{ max}(v_r^2,0) \text{ sin}\theta \text{ }d\theta, \label{eq15}
\end{equation}
and
\begin{equation}
    P_{\rm k}(r)=2\pi r^2 \int_{0^\circ}^{89.9^\circ} \rho \text{ max}(v_r^3,0) \text{ sin}\theta \text{ }d\theta.
\end{equation}

\begin{figure*}[htbp]
\centering
\includegraphics[width=.32\textwidth]{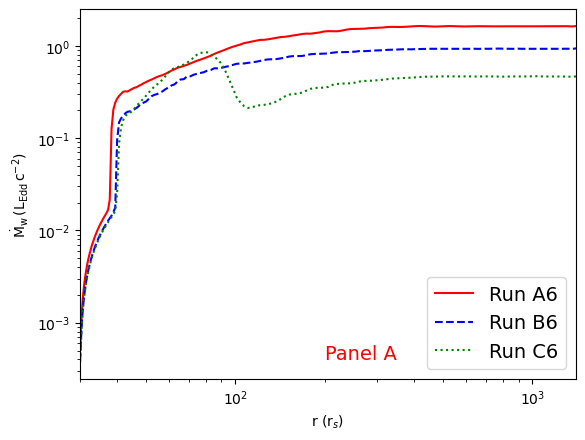}
\includegraphics[width=.32\textwidth]{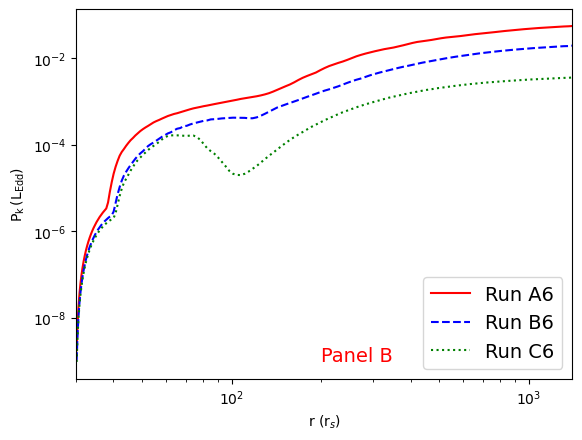}
\includegraphics[width=.32\textwidth]{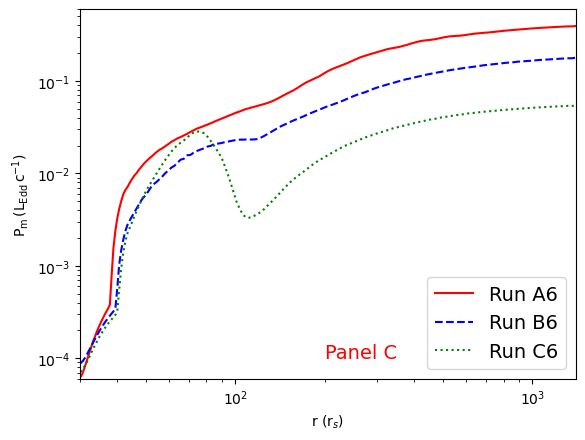}

    \caption{Radial dependence of time-average quantities. Panel (A): the mass outflow rate; panel (B): the kinetic power of winds; panel (C): the momentum flux of winds.}\label{fig.3}
\end{figure*}

Figure \ref{fig.3} shows the time-averaged $\dot{M}_{\rm w}(r)$, $P_{\rm m}(r)$, $P_{\rm k}(r)$. As is shown in figure \ref{fig.3},  $\dot{M}_{\rm w}(r)$, $P_{\rm m}(r)$, $P_{\rm k}(r)$ rapidly increase inside $\sim$40 $r_s$, and then $\dot{M}_{\rm w}(r)$ almost stays constant outside $\sim$100 $r_s$. This implies that winds originates from the disc within 100 $r_{\rm s}$ and most of the material in the wind comes from discs within 40 $r_{\rm s}$. $P_{m}(r)$ and $P_k(r)$ slowly increase on the range of $\sim$40--300 $r_s$, and then slightly increase outside $\sim$300 $r_s$, which indicates that the winds will be continuously accelerated within the computation region, and the region within $\sim$300 $r_s$ is an important region.

\begin{figure*}[htbp]
\centering
\includegraphics[width=.37\textwidth]{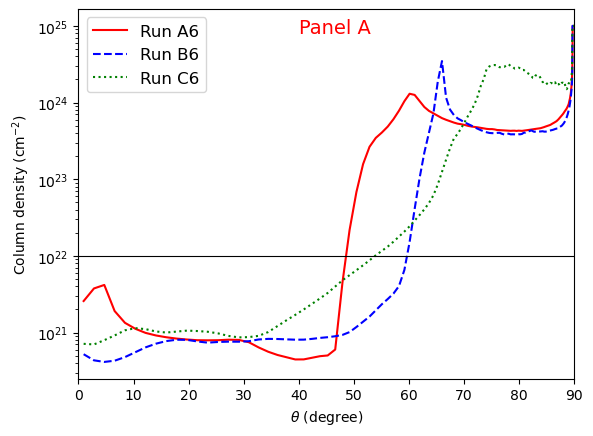}
\includegraphics[width=.37\textwidth]{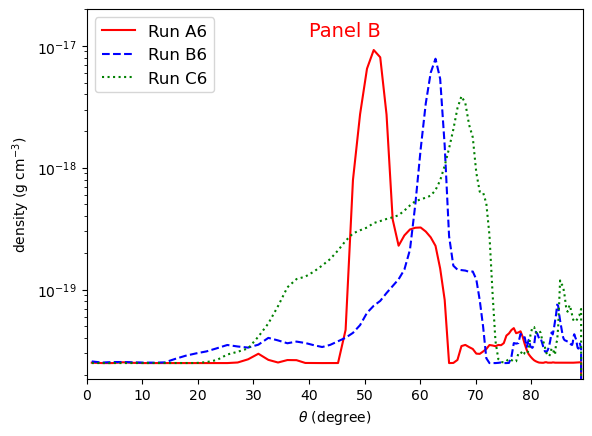}

\includegraphics[width=.37\textwidth]{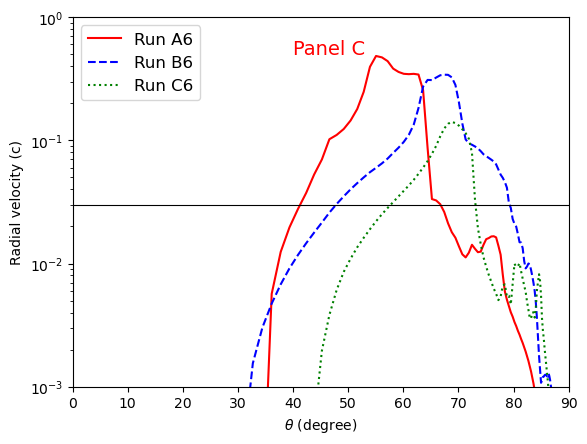}
\includegraphics[width=.375\textwidth]{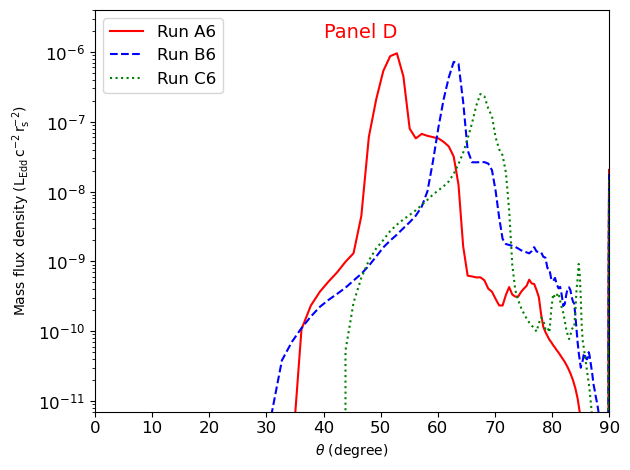}

    \caption{Angular profiles of time-averaged variables. Panel (A): the angular profiles of the column density; Panel (B): the angular profiles of the density at $r=1500 r_{\rm s}$; Panel (C): the angular profiles of the radial velocity at $r=1500 r_{\rm s}$; Panel (D):  the angular profiles of the mass flux density at $r=1500 r_{\rm s}$. In panel (A), the horizontal line means the column density of 10$^{22}$ cm$^{-2}$. In panel (C), the horizontal line means the velocity of 10$^4$ km s$^{-1}$
.}\label{fig.4}
\end{figure*}

Figure \ref{fig.4} shows the angular dependence of wind properties at the outer boundary \textbf{($r=1500r_{\rm s}$)}. In figure \ref{fig.4}, panel (A) shows the angular distribution of the column density ($N_\text{H}=\int_{30r_s}^{1500r_s} \frac{\rho(r,\theta)}{\mu m_{\rm p}}dr$). Panel (A) shows that the column density is higher than 10$^{22}$ cm$^{-2}$ when $\theta>\sim 48^{\circ}$ for run A6, $\theta>\sim60^{\circ}$ for run B6, and $\theta>\sim 53^{\circ}$ for run C6, respectively. At $\theta=\sim 90^{\circ}$, the sudden increase in the column density is caused by the accumulation of material on the disc surface. At $\theta=\sim 80^{\circ}$--90$^{\circ}$, the matter blown off the disc surface significantly contributes to the column density, while its radial velocity remains relatively low. As is shown in panel (A), the matter distribution of run C6 is located closer to the disc surface compared to runs A6 and B6. panel (B) shows the angular distribution of the density at the outer boundary. As is shown in panel (B), the maximum density of winds in runs B6 and C6 is slightly lower than that in run A6. In run B6, the maximum density reaches approximately 85\% of that in run A6, whereas in run C6, it is about 42\% of run A6. When the special relativistic effects are taken into account, the density of winds is reduced. Panel (C) shows the angular distribution of the radial velocity at the outer boundary. For run A6, the radial velocity over $41^{\circ}<\theta<67^{\circ}$ is higher then 10$^4$ km s$^{-1}$. For run B6, the radial velocity higher then 10$^4$ km s$^{-1}$ is located within the angle range of $48^{\circ}<\theta<79^{\circ}$. For run C6, the radial velocity higher then 10$^4$ km s$^{-1}$ is located within the angle range of $58^{\circ}<\theta<73^{\circ}$. For run C6, the maximum radial velocity at the outer boundary reaches 0.14$c$, which is lower than that in runs A6 and B6. In run B6, the maximum radial velocity reaches approximately 70\% of that in run A6, whereas in run C6, it is about 29\% of run A6. Panel (D) shows the angular distribution of the mass flux density at the outer boundary. As is shown in panel (D), the angle range of winds at the outer boundary is about 10$^{\circ}$--20$^{\circ}$. In run B6, the maximum value of the mass flux density reaches approximately 75\% of that in run A6, whereas in run C6, it is about 26\% of run A6. In runs B6 and C6, the reduction in the mass flux density, as compared to run A6, is primarily attributed to a decrease in the radial velocity.

\subsection{Luminosity dependence of wind properties}

\begin{figure*}[htbp]
\centering
\includegraphics[width=.38\textwidth]{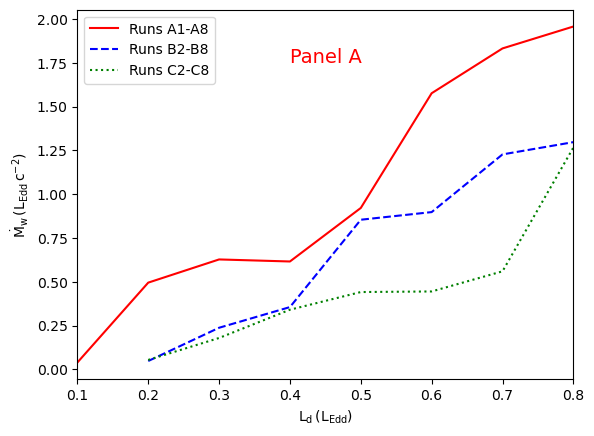}
\includegraphics[width=.37\textwidth]{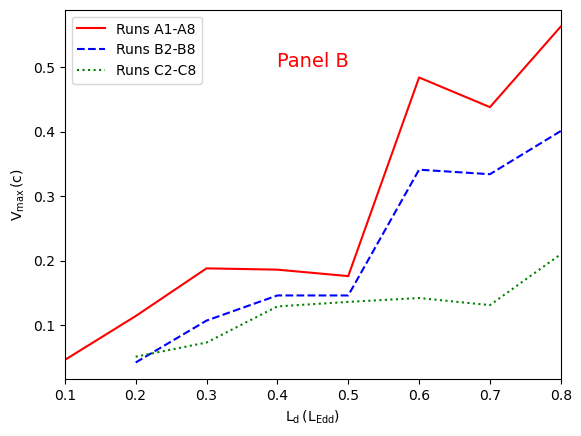}
\includegraphics[width=.38\textwidth]{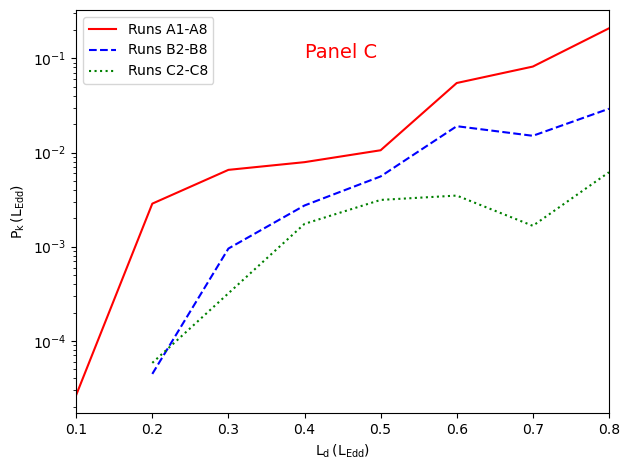}
\includegraphics[width=.37\textwidth]{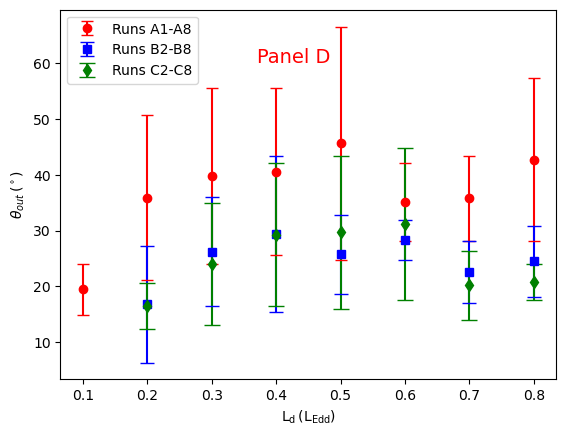}

    \caption{The influence of the disc luminosity on the wind properties at $r=1500 r_{\rm s}$, such as the mass outflow rate ($\dot{M}_{\rm w}$; panel A), the maximum velocity ($v_{\rm max}$; panel B), the kinetic power of winds($P_{\rm k}$; panel C), and the angle of winds (panel D).}\label{fig.5}
\end{figure*}

Figure \ref{fig.5} shows the influences of the disc luminosity on wind properties at the outer boundary \textbf{($r=1500r_{\rm s}$)}, such as the mass outflow rate, the maximum velocity, the kinetic power of winds, and the angle of the winds. When $\epsilon$ = 0.1 (i.e. $L_{\rm d}=0.1L_{\rm Edd}$), the radiation force in run A1 can drive very weak winds, while runs B1 and C1 cannot generate winds that can reach the outer boundary, as is shown in figure \ref{fig.2}. Panel (A) shows that the higher the luminosity, the greater the outflow mass rate. The outflow mass rate in runs C2--C8 is significantly lower than that in runs A2--A8 and is about 10\%--65\% of that in runs A2--A8. Panel (B) shows that the maximum velocity of winds at the outer boundary is greater than 10$^4$ km s$^{-1}$ and reaches the mildly relativistic velocity in all the runs. When the radiation-drag effect is considered in runs C2--C8, the maximum velocity of winds is significantly reduced, compared with runs A2--A8. In runs C2--C8, when $\epsilon$ $<$ 0.4 (i.e. $L_{\rm d}<0.4L_{\rm Edd}$) the maximum velocity is lower than 0.1$c$; when 0.4$\leq\epsilon$ $<$ 0.8 (i.e. $0.4L_{\rm Edd}\leq L_{\rm d}<0.8L_{\rm Edd}$) the maximum velocity is not higher than 0.2 $c$. For run C8, the maximum velocity reaches 0.21 $c$. According to equation (6), as the speed of the winds increases the radiation-drag term becomes important. Comparing runs B6--B8 with runs C6--C8 (also see Table \ref{tab2}), the wind maximum speed is substantially reduced when the radiation-drag effect is taken into account. Therefore, the higher the wind speed, the more significant the radiation-drag effect. Panel (D) shows that the incline angle of winds in runs B2--B8 and C2--C8 is smaller than that in A2--A8, indicating that the winds in runs B2--B8 and C2--C8 are closer to the disc surface than in A2--A8.

\section{Application to UFOs}

\begin{figure*}[htbp]
\centering
\includegraphics[width=.37\textwidth]{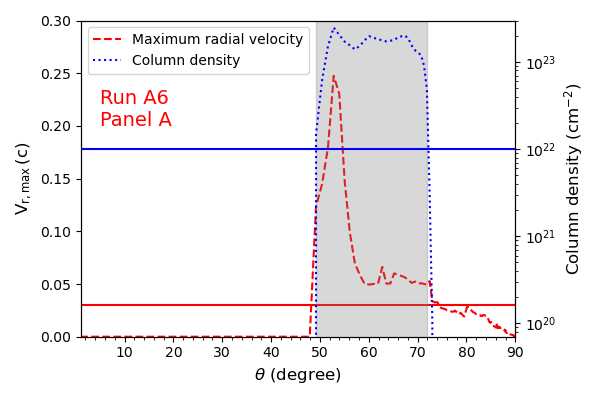}
\includegraphics[width=.37\textwidth]{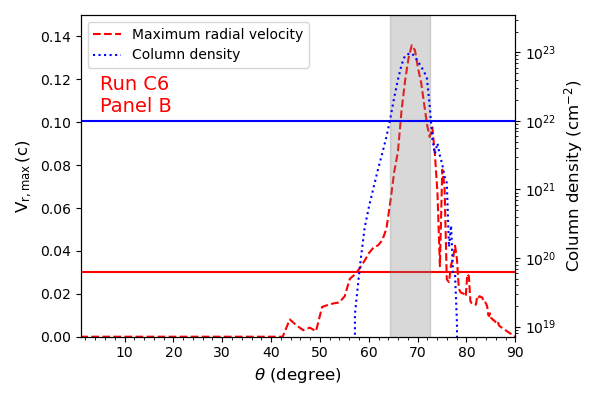}
    \caption{Angular profile of the column density and the maximum radial velocity in runs A6 (left panel) and C6 (right panel). The doted blue line means the column density of the matter with $3\leq\text{log}(\xi/(\text{erg s}^{-1} \text{cm}))\leq6$ and $v_{\rm r}\geq10^4$ km s$^{-1}$ and the dashed red line means the maximum radial velocity of the matter with $3\leq\text{log}(\xi/(\text{erg s}^{-1} \text{cm}))\leq6$. The red solid line and the blue solid line show the line of $v_{\rm r, max}=10^4$ km s$^{-1}$ and $N_{\rm H}=10^{22}$ cm$^{-2}$, respectively. The shaded areas indicate where UFOs could form.}
    \label{fig.6}
\end{figure*}

\cite{2010A&A...521A..57T} first define UFOs as the highly ionized absorbers with a velocity above 10$^4$ km s$^{-1}$. The ionization parameter ($\xi$) and column density of UFOs are distributed in the range of $\text{log}(\xi/(\text{erg s}^{-1} \text{cm})) \sim3-6$ and  $\text{log}(N_{\rm H}/\text{cm}^{-2})\sim22-24$, respectively \citep{2010A&A...521A..57T}. To investigate whether UFOs could be detected from our models or not, we picked out the matter with $3\leq\text{log}(\xi/(\text{erg s}^{-1} \text{cm}))\leq6$ and $v_{\rm r}\geq10^4$ km s$^{-1}$ from each snapshot, then calculated the column density of the matter, and finally time-averaged the column density. We calculated the angular distribution of the column density for runs A6 and C6 and plot it in figure \ref{fig.6}. Figure \ref{fig.6} shows the angular profile of the column density of the matter with $3\leq\text{log}(\xi/(\text{erg s}^{-1} \text{cm}))\leq6$ and $v_{\rm r}\geq10^4$ km s$^{-1}$ and also shows the angular profile of the maximum radial velocity of the matter with $3\leq\text{log}(\xi/(\text{erg s}^{-1} \text{cm}))\leq6$. At a different angle position, the maximum radial velocity may occur at a different radius. The shaded areas in the figure \ref{fig.6} indicate where UFOs could form. For run A6, the column density is higher than 10$^{22}$ cm$^{-2}$ over $\sim$50$^{\circ}$--72$^{\circ}$, where the UFOs could form and the maximum radial velocity of UFOs is about 0.25$c$. For run C6, the UFOs could form over the range of $\sim$64$^{\circ}$--73$^{\circ}$ and the maximum radial velocity of UFOs is about 0.14$c$. Therefore, the special relativistic effects are taken into account, the angular range of UFOs formation is reduced and their maximum radial velocity is also reduced.

The UFOs are detected not only in a substantial proportion of radio-quiet AGNs but also in a limited sample of radio-loud AGNs \citep{2012MNRAS.422L...1T,2014MNRAS.443.2154T}. In radio-loud AGNs, the UFOs are detected at inclination angles of $\sim$10$^{\circ}$--70$^{\circ}$ relative to jets \citep{2014MNRAS.443.2154T}. According to our simulations, we do not expect to observe the line-force-driven UFOs in low-inclination galaxies, such as type-1 Seyfert galaxies. Therefore, there is a significant discrepancy between the UFOs in radio-loud AGNs and the line-force-driven UFOs. To understand the UFOs in radio-loud AGNs, a magnetic-driving model should be suggested \citep{2014ApJ...780..120F,2018ApJ...864L..27F,2021ApJ...922..262Y}. \cite{2021ApJ...922..262Y} suggests that disc winds, driven by both the line force and the magnetic field, can be used to understand the UFOs in radio-loud AGNs. In \cite{2021ApJ...922..262Y} simulations, the UFOs are observed in the angular range of $\sim$10$^{\circ}$--75$^{\circ}$ relative to jets. The simulations in the present work do not take into account the effect of magnetic field. When the effect of the magnetic field is further incorporated into our simulations, the angular range of visibility of the predicted UFOs is expected to increase. In the future, it will be necessary to study the influence of special relativity effects on the winds driven by the magnetic field and line force together. We also note that the special relativistic effects are essentially negligible in pure magnetically driven winds.

\begin{figure*}[htbp]
\centering
\includegraphics[width=.3\textwidth]{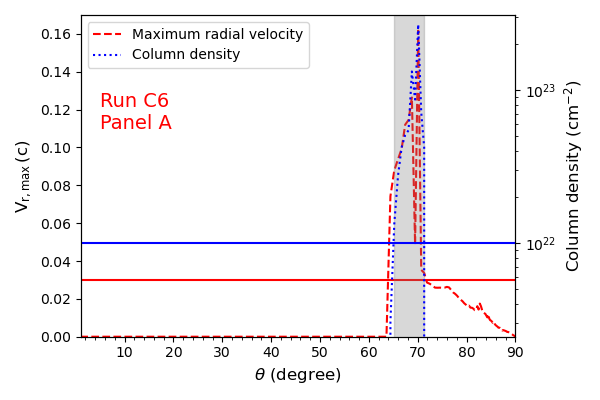}
\includegraphics[width=.3\textwidth]{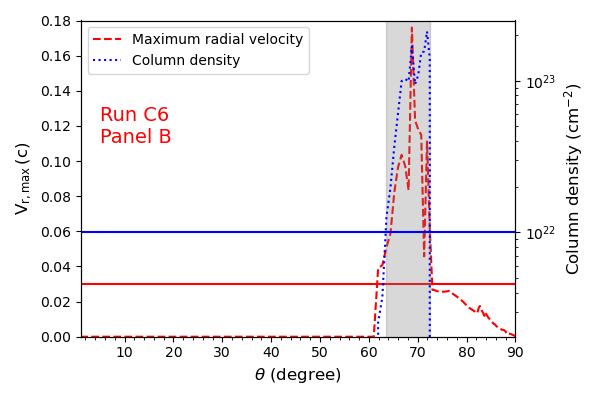}
\includegraphics[width=.3\textwidth]{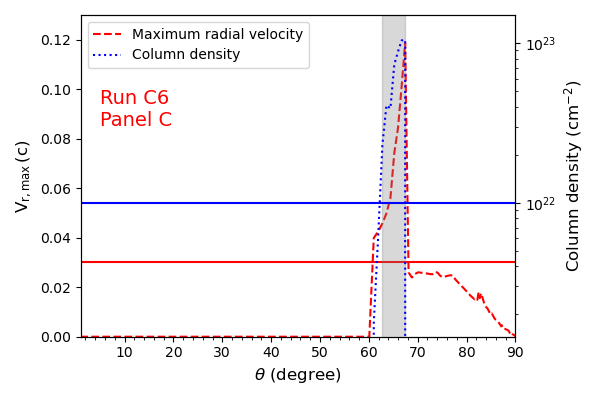}
    \caption{Angular profile of the column density and the maximum radial velocity in runs C6 at $t=1.0$ (panel A), $1.25$ (panel B), and 1.5 $T_{\rm orb}$ (panel C). The meanings of the various lines in the figure are the same as those in figure \ref{fig.6}. The shaded areas indicate where UFOs could form.}
    \label{fig.7}
\end{figure*}

The UFOs are also known to be highly variable with an unknown cadence. To show the variability of UFOs, we present figure \ref{fig.7}, which shows the angular profile of the column density and the maximum radial velocity of UFOs in runs C6 at three time points. The shaded areas in figure \ref{fig.7} indicates where UFOs could form. From figure \ref{fig.7}, we can see that there is slight change in the angle position of formation of the UFOs. This implies that the observed UFOs in our simulations slightly swing. The angular profile of the column density and the maximum radial velocity of UFOs exhibit subtle variations over a short timescale. This can result in the observational features of the theoretical UFOs remaining practically unchanged. Based on the current numerical simulations, a conjecture is that the changes in the observational UFOs may originate from other factors. For example, outflows may induce changes in the accretion flows, which in turn could lead to variation in the outflows themselves \citep{2020MNRAS.494.3616N}.Alternatively, changes occur in the X-ray radiation due to scattering, absorption, and re-emission. The local X-ray radiation determines the gas ionization state and subsequently affects the strength of line force. In order to study the time dependence of the winds launched from the AGN disc, \cite{2024MNRAS.530.5143D} treated the time-dependent radiation transfer of the X-ray radiation field in AGN winds by directly solving the grey (frequency-averaged) time-dependent radiation transport equation. They found that an episodic wind can form.

We also note that the angular range of the formation of UFOs corresponds to a Compton-thick regime, in which scattering can be the dominant emission mechanism. In that case, a significant number of scattered X-ray photons escape from the sides of UFOs, while a smaller quantity of scattered photons could fill in the absorption feature of UFOs. Therefore, a direct consequence would be the dilution of UFO features, i.e. the UFOs absorption line becomes shallower or weaker. The quantitative calculation of the UFO features needs to implement radiative transfer. Based on the post-processing approach, it is feasible to employ the Monte Carlo method to calculate the UFO features. However, this calculation is beyond the scope of the current work.

\section{Conclusions} \label{sec4}

In AGNs, when the disc winds driven by line forces have mildly relativistic velocities, the special relativistic effects become non-negligible. The special relativistic effects are reflected as follows: 1) the influence of the disc rotation on the radiation field, and 2) the radiation-drag effect. We have performed a series of numerical simulations to investigate the influence of the special relativistic effects on the line-force-driven winds. Compared with the case without special relativistic effects, when special relativistic effects are considered in our simulations, the kinetic power of the line-force-driven winds is significantly reduced. Here, we summarise the details of our conclusions as follows:

(1) The special relativistic effects enhance the minimum disc luminosity required to launch line-force-driven winds, raising the luminosity threshold. When the special relativistic effects are not considered, the disc with $L_{\rm d}\geq0.1$ $L_{\rm Edd}$ can produce winds that escape from the gravity of the BH. However, when the influence of the disc rotation on the radiation field is considered, the disc with $L_{\rm d}\geq0.2$ $L_{\rm Edd}$ can produce the winds that escape from the gravity of the BH. The disc of $L_{\rm d}=0.1$ $L_{\rm Edd}$ cannot produce winds that reach an infinite distance.

(2) The influence of the disc rotation on the radiation field causes the line-force-driven winds to be closer to the disc surface and weakens the winds. The maximum speed of winds in the runs of $L_{\rm d}>0.2$ $L_{\rm Edd}$ is reduced by $\sim$ 20 percent--70 percent when the radiation-drag effect is taken into account. The radiation-drag effect further weakens the winds. The higher the wind speed, the more significant the radiation-drag effect. For example, compared with the case without the radiation-drag effect, the maximum speed of winds in the runs of $L_{\rm d}>0.5$ $L_{\rm Edd}$ is reduced by $\sim$ 60 percent--70 percent when the radiation-drag effect is considered.

(3) The mass outflow rate is significantly reduced due to the special relativistic effects.

\begin{acknowledgements}
This work is supported by Chongqing Natural Science Foundation (grant CSTB2023NSCQ-MSX0093) and the Natural Science Foundation of China (grant 12347101). D. Bu is supported by the Natural Science Foundation of China (grant 12173065, 12133008, 12192220, 12192223).
\end{acknowledgements}

\bibliographystyle{aa}
\bibliography{reference}

\appendix
\onecolumn
\section{Calculation of the line force}\label{appendix}

\begin{figure*}[h]
\centering
\includegraphics[width=0.5\textwidth]{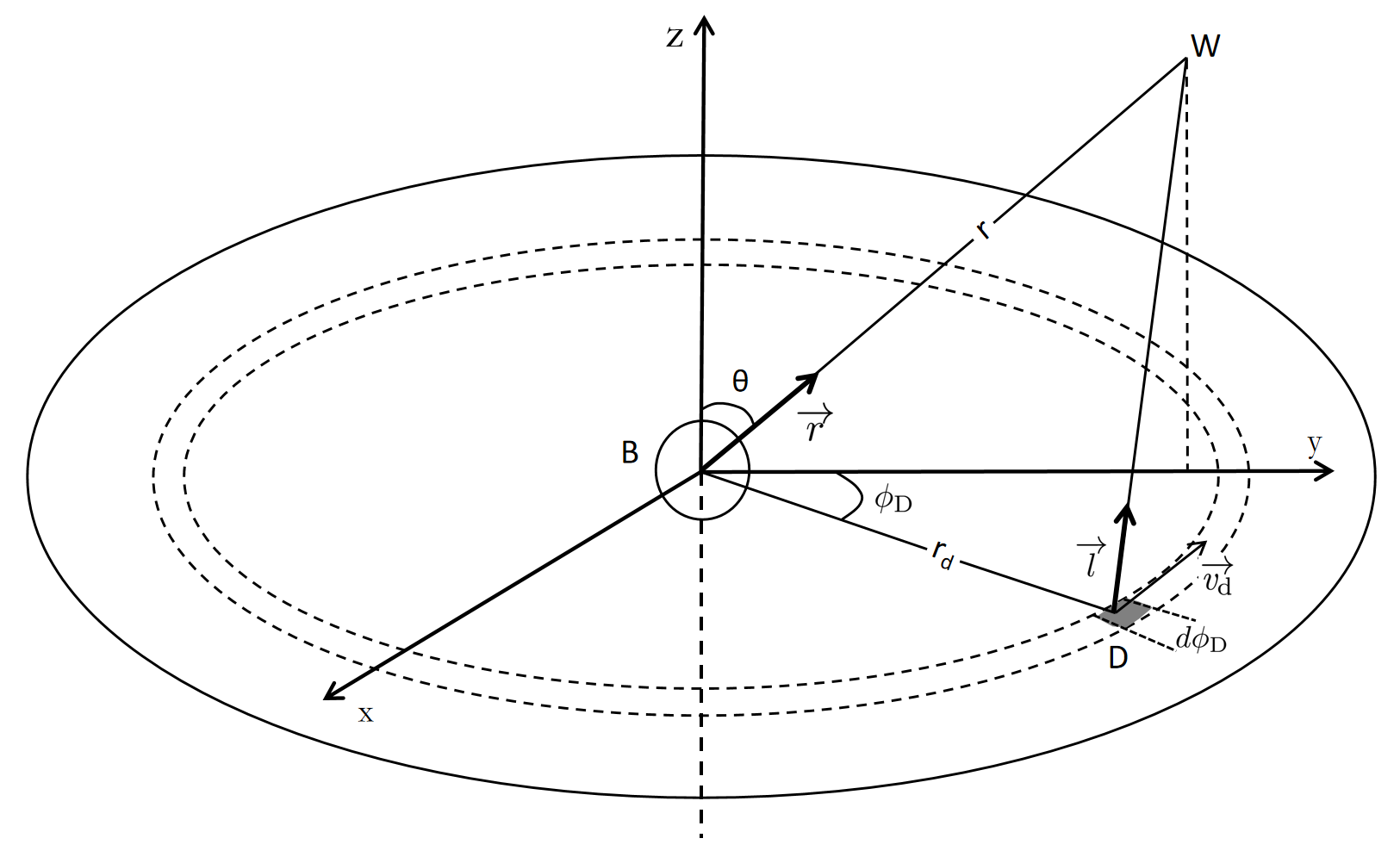}
    \caption{Schematic diagram for calculating the radiation field.}\label{figA1}
\end{figure*}
 Here, we give the calculation of the line force (${g}_{\textit{i}}^{\rm{rad},\textit{l}}$) in equation (\ref{radforce}). The line force exerting on a unit mass is written as
\begin{equation}
    {g}_{\textit{i}}^{\rm{rad},\textit{l}} = \frac{e^{-\tau_{\rm d}}\sigma_{\text{e}}}{\textit{c}}  (f_{1,i}- v_{i} f_2- \sum_{j}v_{j} f_{3,ij}),
\end{equation}\label{eqA1}
where $f_{1,i}$, $f_2$, and $f_{3,ij}$ is defined as:
\begin{equation}
f_{1,i} = \oint_{\Omega}\frac{\mathcal{M} I_{\rm d} l_{i}}{(1-\frac{\overrightarrow{\textit{v}_{\rm d}}\cdot\overrightarrow{\textit{l}}}{c})^4} d\Omega,
\label{eqA2}
\end{equation}
\begin{equation}
f_2 =\frac{1}{c}\oint_{\Omega}\frac{\mathcal{M} I_{\rm d}}{(1-\frac{\overrightarrow{\textit{v}_{\rm d}}\cdot\overrightarrow{\textit{l}}}{c})^4} d\Omega,
\end{equation}
and
\begin{equation}
f_{3,ij}=\frac{1}{c}\oint_{\Omega}\frac{\mathcal{M} I_{\rm d} l_{i} l_{j}}{(1-\frac{\overrightarrow{\textit{v}_{\rm d}}\cdot\overrightarrow{\textit{l}}}{c})^4} d\Omega.
\label{eqA4}
\end{equation}
In equation (\ref{eqA2})--(\ref{eqA4}), $\overrightarrow{\textit{l}}$ and $\overrightarrow{\textit{v}_{\rm{d}}}$ are the direction-cosine vector pointing toward the rest observer (point W) from the disc surface (point D) and the disc velocity, respectively. In spherical co-ordinates, we are given by
\begin{equation}
    \overrightarrow{l}=(\frac{r-r_{\rm d} \text{sin}\theta \text{cos}\phi_{\rm D}} {d_{\rm D}},\frac{-r_{\rm d} \text{cos}\theta \text{cos}\phi_{\rm D}}{d_{\rm D}},\frac{r_{\rm d}\text{sin}\phi_{\rm D}}{d_{\rm D}}),
\end{equation}
and
\begin{equation}
\overrightarrow{\textit{v}_{\rm d}} = (v_{\rm{k}} \text{sin}\theta \text{sin}\phi_{\rm D}, v_{\rm{k}} \text{cos}\theta \text{sin}\phi_{\rm D}, v_{\rm{k}} \text{cos}\phi_{\rm D}),
\end{equation}
where $d_{\rm D}=\left | \rm DW \right |=(r_{\rm d}^2+r^2-2rr_{\rm d}\text{sin}\theta \text{cos}\phi_{\rm D})^{\frac{1}{2}}$ (see figure \ref{figA1}).

Based on the Sobolev approximation, the line-force multiplier ($\mathcal{M}(t)$) in equation (\ref{eqA2})--(\ref{eqA4}) is a function of the optical depth parameter $t = \frac{\sigma_e \rho v_{\rm th}}{\left | dv_l/dl \right|}$, where $v_{\rm th}$ and $\frac{dv_{l}}{dl}$ are the thermal velocity and the velocity gradient along the line of sight, respectively. We set $v_{\rm th}=20 \rm {km/s}$, which corresponds to the gas temperature of 2.5 $\times 10^4\rm{K}$. We employ the CAK approach \citep[][CAK]{1975ApJ...195..157C} modified by \citet[CAK]{1988ApJ...335..914O} to calculate the line-force multiplier, that is given by
\begin{equation}
    \mathcal{M}(t) = k t^{-\alpha} \left\lbrack \frac{(1+\tau_{\rm{max}})^{1-\alpha}-1}{ \tau_{\rm{max}}^{1-\alpha}} \right\rbrack,
\end{equation}
where $\alpha$ is the ratio of optically thick to optically thin spectral lines, $k$ is a parameter proportional to the total number of lines, $\tau_{\rm max}=t\eta_{\rm{max}}$ and $\eta_{\rm{max}}$ is a parameter that determines the maximum value of the line force multiplier ($\mathcal{M}_{\rm{max}}$). In all our simulations, we set $\alpha=0.6$. According to \cite{1990ApJ...365..321S}, $t$ and $\eta_{\rm{max}}$ are respectively give by
\begin{equation}
    k = 0.03 +0.385 \rm{exp}(-1.4 \xi^{0.6})
\end{equation}
and
\begin{equation}
    \text{log}_{10} \eta_{\rm {max}} = \left\{
\begin{aligned}
6.9 \rm{exp}(0.16\xi^{0.4}) \ \ \ \ \ \ \ \ \ \ \ \ \ for \ \ \ \ \ \  log_{10}(\xi) \le 0.5 \\
9.1 \rm{exp}(-7.96 \times 10^{-3} \xi)\ \ \ \ for \ \ \ \ \ \ \ \ log_{10}(\xi) \textgreater 0.5
\end{aligned}
\right.\ \ .
\label{eqA9}
\end{equation}
For $dv_l/dl$, we can write as \citep{1978ApJ...219..654R}
\begin{equation}
    \frac{dv_{l}}{dl}=Q\equiv\sum_{i,j}\frac{1}{2}(\frac{\partial v_{i}}{\partial r_{j}}+\frac{\partial v_{j}}{\partial r_{i}})l_{i}l_{j},
\end{equation}
where $r_{i}$ is the components of the position vector ($\bf{r}$) and $l_{i}$ is the components of the unit vector ($\hat{\bf{l}}$) of the direction D toward W (see figure \ref{figA1}), respectively. We assume that $dv_l/dl$ is dominated by the gradient of the velocity along the vertical direction and then $dv_l/d_l$ is given by
\begin{equation}
    \frac{dv_l}{dl} = \frac{dv_{\rm z}}{d\rm z}n_{\rm z}^2 =\left \lbrack \text {cos}^2\theta \frac{\partial v_r}{\partial r} + \text{sin}^2 \theta \frac{1}{r} (v_r + \frac{\partial v_\theta}{\partial \theta}) +\text{sin}\theta \text{cos}\theta (\frac{v_\theta}{r} - \frac{\partial v_\theta}{\partial r} -\frac{\partial v_r}{r\partial \theta}) \right \rbrack n_{\rm z}^2,
\end{equation}
where $n_{\rm z}=\frac{r\text{cos}\theta }{d_{\rm D}}$. For the element D, its area $dS=r_{\rm d} dr_{\rm d} d\phi_{\rm D}$. The solid angle subtended by the element D towards the observer standing at the point W is $d\Omega=\frac{r\text{cos}\theta dS}{d_{\rm D}^3}$. Therefore, $f_{1,i}$ can be written
\begin{equation}
    f_{1,i} =  \frac{9 \epsilon L_{\rm{Edd}} r_{\rm s}}{4 \pi^2} k(\sigma_{\rm{e}} \rho v_{th} |\frac{dv_{\rm {z}}}{d{\rm{z}}}|^{-1})^{-\alpha} \int_{3r_{\text{s}}}^{1500r_{\text{s}}} \int_{0}^{2\pi} \frac{r \text{cos}_\theta}{ r_{\rm d}^2 d_{\rm D}^3} \left( 1-\sqrt{\frac{3r_{\rm s}}{r_{\rm d}}} \ \right) \left\lbrack \frac{(1+\tau_{\rm{max}})^{(1-\alpha)}-1}{\tau_{\rm {max}}}\right\rbrack (\frac{r\text{cos}_\theta}{d_{\rm{D}}})^{2\alpha} \frac{l_i}{(1-\frac{\overrightarrow{\textit{v}_{\rm d}}\cdot\overrightarrow{\textit{l}}}{c})^4} dr_{\rm d} d\phi_{\rm D},\label{eqA12}
\end{equation}
In the above integrand, the $\tau_{\rm{max}}$ term is time-dependent and then strictly this term cannot be removed from the integrand. For removing the $\tau_{\rm{max}}$ term from the integrand, we refer to \cite{1998MNRAS.295..595P} and then make an approximation that $n_{\rm z}=1$ for $\tau_{\rm{max}}$. We introduce a new variable $\tau_{\rm{max}}^{\prime}=\sigma_{\rm e}\rho v_{th}|dv_{\rm z}/d\rm{z}|^{-1} \eta_{\rm{max}}$ and rewrite equation (\ref{eqA12}) as
\begin{equation}
    f_{1,i} =  \frac{9 \epsilon L_{\rm{Edd}} r_{\rm s}}{4 \pi^2} k(\sigma_{\rm{e}} \rho v_{th} |\frac{dv_{\rm {z}}}{d{\rm{z}}}|^{-1})^{-\alpha}\left\lbrack \frac{(1+\tau_{\rm{max}}^{\prime})^{(1-\alpha)}-1}{\tau_{\rm {max}}^{\prime}}\right\rbrack \int_{3r_{\text{s}}}^{1500r_{\text{s}}} \int_{0}^{2\pi} \frac{r \text{cos}_\theta}{ r_{\rm d}^2 d_{\rm D}^3} \left( 1-\sqrt{\frac{3r_{\rm s}}{r_{\rm d}}} \ \right) (\frac{r\text{cos}_\theta}{d_{\rm{D}}})^{2\alpha} \frac{l_i}{(1-\frac{\overrightarrow{\textit{v}_{\rm d}}\cdot\overrightarrow{\textit{l}}}{c})^4} dr_{\rm d} d\phi_{\rm D}.
\end{equation}
Similarly, $f_2$ and $f_{3,ij}$ can be expressed as
\begin{equation}
    f_{2} =  \frac{9 \epsilon L_{\rm{Edd}} r_{\rm s}}{4 \pi^2 c} k(\sigma_{\rm{e}} \rho v_{th} |\frac{dv_{\rm {z}}}{d{\rm{z}}}|^{-1})^{-\alpha}\left\lbrack \frac{(1+\tau_{\rm{max}}^{\prime})^{(1-\alpha)}-1}{\tau_{\rm {max}}^{\prime}}\right\rbrack \int_{3r_{\text{s}}}^{1500r_{\text{s}}} \int_{0}^{2\pi} \frac{r \text{cos}_\theta}{ r_{\rm d}^2 d_{\rm D}^3} \left( 1-\sqrt{\frac{3r_{\rm s}}{r_{\rm d}}} \ \right) (\frac{r\text{cos}_\theta}{d_{\rm{D}}})^{2\alpha} \frac{1}{(1-\frac{\overrightarrow{\textit{v}_{\rm d}}\cdot\overrightarrow{\textit{l}}}{c})^4} dr_{\rm d} d\phi_{\rm D},
\end{equation}
and
\begin{equation}
    f_{3,ij} =  \frac{9 \epsilon L_{\rm{Edd}} r_{\rm s}}{4 \pi^2 c} k(\sigma_{\rm{e}} \rho v_{th} |\frac{dv_{\rm {z}}}{d{\rm{z}}}|^{-1})^{-\alpha}\left\lbrack \frac{(1+\tau_{\rm{max}}^{\prime})^{(1-\alpha)}-1}{\tau_{\rm {max}}^{\prime}}\right\rbrack \int_{3r_{\text{s}}}^{1500r_{\text{s}}} \int_{0}^{2\pi} \frac{r \text{cos}_\theta}{ r_{\rm d}^2 d_{\rm D}^3} \left( 1-\sqrt{\frac{3r_{\rm s}}{r_{\rm d}}} \ \right) (\frac{r\text{cos}_\theta}{d_{\rm{D}}})^{2\alpha} \frac{l_i l_j}{(1-\frac{\overrightarrow{\textit{v}_{\rm d}}\cdot\overrightarrow{\textit{l}}}{c})^4} dr_{\rm d} d\phi_{\rm D}.
\end{equation}

\end{document}